%% file: main.tex
\documentclass[journal]{IEEEtran}

\usepackage{amssymb,amsmath,bm}
\usepackage{cite}
\usepackage{commath}
\usepackage{latexsym}
\usepackage{tabularx}
\usepackage[]{graphicx}
\usepackage[caption=false]{subfig}
\usepackage{mathtools}
\usepackage{siunitx}

\usepackage{listings}
\usepackage{tikz}
\usepackage{breqn}
\usepackage{stfloats}
\usepackage{comment}
\usepackage{fontawesome5}
\usepackage{longtable}
\usepackage[colorlinks=true,linkcolor=black,anchorcolor=black,citecolor=black,filecolor=black,menucolor=black,runcolor=black,urlcolor=black]{hyperref}
\usepackage{ragged2e}
\usepackage[]{array}
\usepackage{multirow}
\usepackage{soul}
\usepackage{booktabs}

\usepackage{textcomp}
\usepackage{enumitem}

\usepackage{xcolor}
\usepackage[linesnumbered,ruled,vlined]{algorithm2e}
\usepackage{algpseudocode}

\DeclareMathOperator*{\argmin}{arg\,min}

\IEEEoverridecommandlockouts

\makeatletter

\newcolumntype{L}[1]{>{\raggedright\let\newline\\\arraybackslash\hspace{0pt}}m{#1}}
\newcolumntype{C}[1]{>{\centering\let\newline\\\arraybackslash\hspace{0pt}}m{#1}}
\newcolumntype{R}[1]{>{\raggedleft\let\newline\\\arraybackslash\hspace{0pt}}m{#1}}

\makeatletter
\newcommand\remembertext[2]{
  \immediate\write\@auxout{\unexpanded{\global\long\@namedef{mytext@#1}{#2}}}%
   {%
   \hypersetup{linkcolor=black,anchorcolor=black,citecolor=black,filecolor=black,menucolor=black,runcolor=black,urlcolor=black}%
   \color{black} #2%
   \hypersetup{linkcolor=black,anchorcolor=black,citecolor=black,filecolor=black,menucolor=black,runcolor=black,urlcolor=black}%
   }%
}
\newcommand\recalltext[1]{%
  \ifcsname mytext@#1\endcsname
    {%
    \hypersetup{linkcolor=black,anchorcolor=black,citecolor=black,filecolor=black,menucolor=black,runcolor=black,urlcolor=black}%
    \color{black}\@nameuse{mytext@#1}%
    \hypersetup{linkcolor=black,anchorcolor=black,citecolor=black,filecolor=black,menucolor=black,runcolor=black,urlcolor=black}%
    }%
  \else
    ``???''
  \fi
}
\makeatother

\input{acronyms}

\begin{document}


\title{Algorithm-Supervised Millimeter Wave Indoor\\ Localization using Tiny Neural Networks}


%





\author{
Anish~Shastri,~\IEEEmembership{Graduate~Student~Member,~IEEE},
Steve~Blandino,~
Camillo~Gentile,~\IEEEmembership{Member,~IEEE},
Chiehping~Lai,~\IEEEmembership{Member,~IEEE},
Paolo~Casari,~\IEEEmembership{Senior~Member,~IEEE}\vspace{-2mm}
\thanks{Manuscript received xxxx xx, xxxx \ldots}%
\thanks{A.~Shastri and P.~Casari are with DISI, University of Trento, 38123 Povo (TN), Italy (emails: anish.shastri@unitn.it, paolo.casari@unitn.it).}
\thanks{S.~Blandino is research associate with the
National Institute of Standards and Technology (NIST), Gaithersburg, MD  20899 USA, and contractor with Prometheus Computing LLC, Bethesda, MD 20814 USA (email: steve.blandino@nist.gov).}
\thanks{C.~Gentile (email: camillo.gentile@nist.gov) and C.~Lai (e-mail: chiehping.lai@nist.gov) are with the Radio Access and Propagation Metrology Group, National Institute of Standards and Technology, Gaithersburg, MD 20899 USA.}%
\thanks{This work received support from the European Commission's Horizon 2020 Framework Programme under the Marie Sk{\l}odowska-Curie Action MINTS (GA no.~861222), and was partially supported by the European Union under the Italian National Recovery and Resilience Plan (NRRP) of NextGenerationEU, partnership on ``Telecommunications of the Future'' (PE00000001 - program ``RESTART'').}
\thanks{U.S. Government work, not subject to U.S. Copyright.}
}
%




\maketitle
\begin{abstract}
The quasi-optical propagation of \ac{mmw} signals enables high-accuracy localization algorithms that employ geometric approaches or machine learning models. However, most algorithms require information on the indoor environment, may entail the collection of large training datasets, or bear an infeasible computational burden for \ac{cots} devices. In this work, we propose to use tiny \acp{nn} to learn the relationship between \ac{adoa} measurements and locations of a receiver in an indoor environment. To relieve training data collection efforts, we resort to an algorithm-supervised approach by bootstrapping the training of our neural network through location estimates obtained from a state-of-the-art localization algorithm. We evaluate our scheme via \ac{mmw} measurements from indoor 60-GHz double-directional channel sounding. We process the measurements to yield dominant multipath components, use the corresponding angles to compute \ac{adoa} values, and finally obtain location fixes. Results show that the tiny \ac{nn} achieves sub-meter errors in 74\% of the cases, thus performing as good as or even better than the state-of-the-art algorithm, with significantly lower computational complexity.
\end{abstract}


%
\IEEEpeerreviewmaketitle

\glsresetall

\section{Introduction}  \label{sec:intro}



\Ac{mmw} signals are characterized by wide bandwidth and are transmitted via directional arrays. These characteristics are ideal to provide accurate range and angle information, which are essential for precise localization~\cite{fettweis_loc_feature_mmwave_IWCMC_2016}.
In fact, \ac{mmw} signals propagate quasi-optically: they reflect off surfaces and obstacles almost specularly, with minimum scattering and diffraction, and are short-ranged due to small antenna aperture and high atmospheric attenuation. 
The typically sparse channels that result include a dominant \ac{los} path and a few \ac{nlos} paths, which can often be mapped to the scatterer that generated them to achieve high-accuracy localization and environment mapping~\cite{henk2022radioLocFund}. 

\ac{mmw} signals attenuate over short distances and humans are their most significant blockers indoors~\cite{humanBlockage}. Hence, dense \ac{ap} deployments will be inevitable to provide uniform coverage and support next-generation applications requiring multi-Gbit/s data rates such as \ac{vr}, 4K video streaming, and indoor robot navigation~\cite{mmwaveComst2017Hanzo, comst2023anish}. 
In this scenario, accurate localization 
helps optimize the performance of \ac{mmw} networks by designing optimized beam training protocols~\cite{ubiqPefkianakis}, or prevent blockage events by triggering timely handovers~\cite{leap2019palacios}. Moreover, device-centric algorithms lead to better scalability in dense \ac{mmw} networks and help reduce the network maintenance overhead, increase spatial reuse, and  enable load balancing and \ac{mac}-level communication scheduling~\cite{scalingMmwaveClaudio}.

Localization schemes should cater for the computational constraints of \ac{cots} devices and work with existing \ac{mmw} communication equipment, with no need for extra sensors~\cite{yang2021integratedLocComm,xiao2020overview6G}. 
Yet, current approaches tend to cluster into one of the following categories: either $i)$ they employ large \ac{nn} models to process \ac{mmw} radio features and output a location estimate, or $ii)$ solve complex geometric and optimization problems. In case $i)$, the resulting algorithms may fit \ac{cots} devices, but require an often burdensome training data collection phase. In case $ii)$, the algorithms may make strong assumptions related to environment knowledge (e.g., they may assume to know the absolute orientation of the devices), or their complexity may grow super-linearly with the number of measurements collected. In both cases, the algorithm may become impractical in realistic scenarios.

In this work, we propose to bridge the above gap by means of an algorithm-supervised indoor localization approach. To keep complexity minimal, we resort to a tiny \ac{nn} model that exploits \ac{adoa} information as input features. Such \ac{nn} models have much fewer weights compared to other architectures in the literature, thus being compatible with resource-constrained \ac{cots} hardware. 
To relieve the effort of collecting training data for the \ac{nn}, we initially resort to a bootstrapping localization algorithm based on the same \ac{adoa} data we feed to the \ac{nn}. 
While such algorithm will inevitably make errors when computing location estimates, it will still yield approximate location labels for \ac{adoa} readings, which can work as training data for the tiny \ac{nn}. In any event, the advantage of not having to explicitly collect training data largely offsets the disadvantage of imperfect location labels in any practical scenario.
As each \ac{mmw} device localizes itself, our scheme is \emph{device-centric}, hence more scalable than \ac{ap}-centric approaches~\cite{comst2023anish}.

We choose JADE as the bootstrapping%
    \footnote{In this work, we refer to bootstrapping as the data annotation technique that automatically obtains labels for the training data.}  %
algorithm. JADE is a state-of-the-art geometric approach from the literature~\cite{palacios2017jade} that uses \ac{adoa} information to solve a grid search optimization and two iterative \ac{mmse} problems. It jointly localizes a \ac{mmw} device and all surrounding physical and virtual \acp{ap},%
    \footnote{Virtual \acp{ap} appear as the source of \ac{nlos} paths originating from a physical \ac{ap} and reflecting off an obstacle. 
    }   %
without any prior information about the environment (number of \acp{ap}, room boundaries, etc). 

We evaluate the performance of our approach using channel measurements, recorded with a high-precision 3D double-directional 60-GHz channel sounder throughout four different transmitter locations and one mobile receiver, totaling \textgreater{}10\,000 measurements in a fully furnished work environment. 
The rich output of the sounding process includes numerous \acp{mpc}, which cannot be directly used for localization purposes. Different from previous approaches~\cite{accurate3D2018pefkianakis,anish2022wcnc}, we pre-process and filter such output. There are two reasons why this is necessary. First, the wealth of sensed \acp{mpc} makes it challenging to associate them to ambient scatterers.
This is critical, as wrong associations may jeopardize the localization process.
Second, while the \ac{los} path and first-order reflections yield the majority of the received energy, diffuse paths scattered from rough surfaces 
may also bear non-negligible energy. Therefore, we need to identify them out of the remaining \acp{mpc}. 

\remembertext{contributions}{This paper substantially extends our previous work~\cite{anish2022wcnc}, which proposed shallow \ac{nn} models working in simple, simulated indoor environments. In this paper, we employ \acp{nn} that apply to more realistic scenarios, and validate their performance using real channel measurements. 
Moreover, we provide additional simulation results that show how location errors distribute across an indoor space in different conditions. 

The specific contributions of our work are:\\
\indent a) We propose a tiny, computationally-lightweight \ac{nn} model that estimates the location of a \ac{mmw} device by leveraging \ac{adoa} measurements obtained from estimated multipath components at the receiver. 

b) We reduce the training dataset collection overhead via an algorithm-supervised approach: we train the \ac{nn} model using error-prone location estimates from a bootstrapping geometric scheme that also exploits \acp{adoa}; once trained, we only use the \ac{nn} for localization.

c) We evaluate our approach using a rich \ac{mmw} channel sounding dataset collected at NIST Boulder's campus~\cite{9664350}, which we extensively process. 
In particular, we propose a recursive clustering algorithm based on \ac{dbscan} to distinguish  the dominant \ac{los} \ac{mpc} from first-order-reflected \acp{mpc} and the diffused \acp{mpc}.

d) We complement our evaluation with additional simulation results from a complex-shaped room environment.}

The remainder of this paper reviews indoor \ac{mmw} localization techniques (\S\ref{sec:related}); describes our proposed algorithm (\S\ref{sec:algo}) and the experimental data used to validate it (\S\ref{sec:dataset}); and discusses experimental and simulation results (\S\ref{sec:results}). Finally, \S\ref{sec:concl} concludes the paper.
 
 
\section{Related work}  \label{sec:related}


\subsection{Conventional localization and mapping algorithms}
Location systems based on \ac{mmw} technology employ one or more signal parameters 
to estimate the location of a device~\cite{comst2023anish,indoorLocSurvey2019zafari}.
Schemes relying on \ac{rssi} and/or \ac{snr} information typically infer the distance between the device and the known anchors by inverting indoor path loss models. For instance, in~\cite{rssi2014icc} the authors propose trilateration-based localization using \ac{rssi} measurements for a 60\,GHz IEEE~802.11ad network. 
 
Most of the algorithms based on angle information employ geometric techniques~\cite{comst2023anish}. The simplest schemes triangulate the receiver using \acp{aoa} from multiple transmitters, or compute \acp{adoa} across different \ac{mpc} pairs to localize a device~\cite{olivier2016lightweight,palacios2019single}. 
In~\cite{palacios2017jade}, the authors propose JADE, a scheme that jointly localizes both the devices and the \acp{ap}, and then reconstructs the likely location of reflective surfaces in the environment. The scheme requires multiple iterations before location estimates converge, and needs \ac{adoa} measurements from different locations before to be accurate. New \ac{adoa} measurements help refine previous location estimates. However, the complexity of JADE increases quadratically with the number of measurements, making it too complex to be implemented on resource-constrained \ac{cots} devices.
Blanco \emph{et al.} exploit \ac{aoa} and \ac{tof} measurements from 60-GHz 802.11ad-based
routers as well as sub-6~GHz equipment to trilaterate the location of a client~\cite{blanco2022mobisys}.


The \ac{ap}-centric localization algorithm in~\cite{leap2019palacios} exploits \ac{csi} measurements from a client to infer angle information, and thus the client's location. Location estimates were then employed to enhance handover decisions and beam pattern selection.
A~map-assisted positioning technique to localize a device was proposed in~\cite{kanhere2019map} and evaluated using a 3D ray tracer at 28\,GHz and 73\,GHz. 
Yassin et al.\ worked on localization and mapping schemes based on context inference, obstacle detection and classification using geometric techniques~\cite{Yassin2017SimultaneousCI, Yassin2018GeometricAI, MOSAIC2018slam}.
The scheme in~\cite{clam2018Palacios} uses the beam training procedure to acquire \acp{aoa}, employs \acp{adoa} to localize the \ac{mmw} device and all the  anchors to simultaneously map the environment. The scheme is experimentally evaluated on 60\,GHz \ac{mmw} hardware. 
In~\cite{bielsa2018indoor}, the authors leverage coarse-grained per-beampattern \ac{snr} measurements provided by a modified firmware flashed on multiple 802.11ad-compliant 
routers. 
\Ac{aod} and \ac{snr} were used in~\cite{BeamAoD} to design beam-based midline intersection and beam scaling-based positioning algorithms. These were evaluated using both ray-tracing and WiGig \ac{soc} transceivers. 
%

\vspace{-2mm}
\subsection{Machine learning-based algorithms}
\label{subsec:dl.survey}
Several works employed deep learning to localize \ac{mmw} devices indoors. For example, a multi-layer perceptron regression model maps the \ac{snr} fingerprints to the coordinates of a device in one dimension in~\cite{Vashist2020ml}.
Pajovic \emph{et al.}\ used \ac{rssi} and beam index fingerprints to design probabilistic localization models~\cite{fing2019mitsubishi1} and used spatial beam \acp{snr} in~\cite{fing2019mitsubishi2} to learn a multi-task model for position and orientation classification. The authors in~\cite{deepL2020mitsubishi} and~\cite{wangMERLfingerprintingPart4} proposed ResNet-inspired models for \ac{los} and \ac{nlos} scenarios.
To tackle \ac{nlos} propagation, spatial beam \ac{snr} values were used in~\cite{deepL2020mitsubishi}, and multi-channel beam covariance matrix images in~\cite{wangMERLfingerprintingPart4}. 
In~\cite{wbfps2020Infocom}, the authors used the \ac{knn} algorithm to localize \acp{uav} via indoor RSSI-based fingerprinting, whereas \acp{aoa} fingerprints together with \ac{rssi} measurements at reference locations were used in~\cite{doaLf2017}.

\remembertext{dlSurvey}{Although deep learning schemes yield accurate location estimates, they (a) are not easily scalable, as radio fingerprints change with the number of \acp{ap} as well as the setup of the indoor environment (furniture, presence of people, etc.); (b) require knowledge of the indoor environment; (c) require large training datasets; (d) require fusing multiple \ac{mmw} signal parameters to form rich input feature vectors for the models; (e) rely on models with hundreds of thousands of parameters, which may be infeasible for resource-constrained devices.}

\vspace{-2mm}
\subsection{Summary}

The above survey suggests that experimentally evaluated \ac{mmw} localization schemes tend to exhibit one of the following limitations: (a) Low-complexity geometric techniques such as in~\cite{palacios2019single} necessitate knowledge of a device's orientation and environment map. Furthermore, they are vulnerable to imprecise \ac{aoa} estimates. (b) Geometric schemes that gradually refine location estimates from multiple \ac{aoa} measurements experience super-linearly high computational complexity as the number of collected measurements increases. (c) Deep learning-based techniques rely on fingerprint datasets, which require a significant training data collection burden (particularly in large and challenging environments); moreover, they may necessitate excessive computational resources. (d) Implementing and deploying these algorithms on resource-constrained \ac{cots} devices is challenging.

By way of contrast, our tiny \acp{nn} have much fewer parameters to learn, thus have lower computational complexity and can train faster. We avoid collecting ground truth data for our model by bootstrapping the training process via location estimates from a geometric localization algorithm. 
After training, we switch to the \ac{nn} model, and do not need the bootstrapping algorithm any longer for the same environment. Our models achieve a good accuracy with real 60\,GHz channel measurements, despite training with inherently error-prone labels.


\section{Proposed localization scheme}  \label{sec:algo}

\begin{figure}[t]
\centering
\includegraphics[width=\columnwidth]{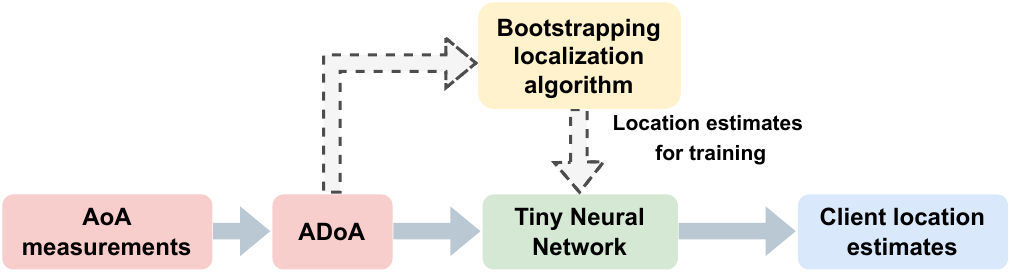}
 \caption{Workflow of our proposed localization scheme.
 }
 \label{fig:nnJadeFlow}
 \vspace{-2mm}
\end{figure}

\subsection{Key idea: algorithm-supervised tiny neural networks} \label{sec:algo.key}

Our key proposition is to localize \ac{mmw} devices via a tiny \ac{nn}, trained to reproduce the behavior of an otherwise complex localization algorithm. Besides the lower computational burden that results, using a bootstrapping algorithm allows us to train the \ac{nn} in a fully algorithm-supervised fashion, with no need for the typically heavy process of collecting training data.

Fig.~\ref{fig:nnJadeFlow} shows the workflow of our proposed localization scheme. \remembertext{syncComment}{We assume that the SNR between the transmitter and receiver is sufficient to achieve over-the-air synchronization.}
We start by computing \acp{adoa} from the set of \acp{mpc} of the \ac{mmw} channel at a given location, by taking the \ac{aoa} of one \ac{mpc} as a reference. 
A tiny \ac{nn} model then learns a non-linear regression function, mapping the input \ac{adoa} values to the 2D coordinates of the receiver. As training is algorithm-supervised via a bootstrapping localization algorithm, the association between \ac{adoa} and the location where these \acp{adoa} are observed will be inherently error-prone. 
Note that classification models, which would return the best-matching location from a training database for each set of measured \acp{adoa}, would be inherently less robust to error-prone training data than regression \ac{nn} models~\cite{anish2022wcnc}, besides requiring a much larger \ac{nn} output stage with inherently coarse location resolution. In fact, algorithm-supervised training with error-prone location labels is advantageous in our case, both from a practical standpoint (because it removes the need for training data collection) and from an accuracy standpoint.

\subsection{Input features of the NN}

\remembertext{input1a}{We employ \ac{adoa} values computed from the \acp{aoa} of the \acp{mpc} of the channel extracted by the receiver, to localize \ac{mmw} devices}.%
    \footnote{\label{fn:delayGap}\remembertext{delayGap}{We neglect the \ac{mpc} delay as the input to both the schemes, as most commercially deployed indoor \ac{mmw} systems such as IEEE 802.11ad/ay \ac{wlan} systems require additional \ac{csi} pre-processing steps in order to retrieve the \ac{mpc} delays, unlike the \acp{aoa}. While there exist standards for estimating the time of flight, e.g., via the \ac{ftm} protocol (IEEE 802.11-REVmc), these methods generally estimate the delay corresponding to the strongest path between the \ac{tx} and the \ac{rx}, which can be either a \ac{los} or \ac{nlos} path.}}  %
\remembertext{input1b}{This removes the compass bias at the receiver, thus making localization problem invariant to \ac{rx} orientation~\cite{olivier2016lightweight}.} Note that this procedure is amenable both to lab-grade devices~\cite{accurate3D2018pefkianakis,clam2018Palacios} and \ac{cots} devices~\cite{bielsa2018indoor,leap2019palacios}.

\begin{figure}[t]
\centering
\includegraphics[width=0.9\columnwidth]{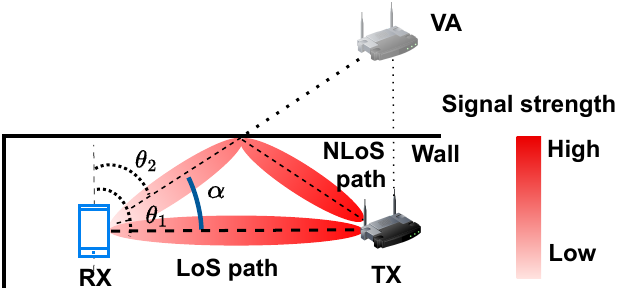}
 \caption{Illustration of virtual anchor and first-order reflections. Here, $\theta_1$ and $\theta_2$ represent the \acp{aoa} of the \ac{los} and the \ac{nlos} paths, and $\alpha = \theta_1 - \theta_2$ represents the \ac{adoa} with respect to the \ac{los} path. Note that first-order reflections have the same geometrical properties (delay and AoA) of direct rays generated from virtual anchors. }
 \label{fig:VAdiag}
\end{figure}

The set of \acp{mpc} that reach the receiver typically includes one \ac{los} path 
and multiple \ac{nlos} paths. 
\ac{nlos} paths can be modeled as emanating from a (virtual) source that corresponds to the mirror image of the physical \ac{tx} with respect to a reflective surface. We refer to these sources as virtual \acp{tx}, and collectively call physical and virtual \acp{tx} \emph{anchors}.
Fig.~\ref{fig:VAdiag} illustrates one \ac{va} for the \ac{tx}, representing the source of the \ac{nlos} path. 



After extracting the \acp{mpc} from each available \ac{tx}, we elect one reference \ac{mpc} and compute \acp{adoa} with respect to the \ac{aoa} of that \ac{mpc}. The reference \ac{mpc} can be the \ac{los} path or a \ac{nlos} path from any transmitter.
Once we collect \ac{aoa} measurements from all $N_a$ anchors, we obtain $N_a - 1$ \ac{adoa} values, which are employed as the input features to our \ac{nn}.

Although the \ac{los} and specular \acp{mpc} account for most of the channel energy, diffuse \acp{mpc} arising from rough surface scattering may appear, especially at \acp{mmw} frequencies. The number of \acp{mpc} that reach the \ac{rx} is thus typically larger than the total number of anchors. 
Hence, \ac{mmw} propagation in indoor environments is often modelled through clusters of rays~\cite{mmwaveComst2017Hanzo, comst2023anish, 8581503}. 
Similarly, we propose to cluster the \acp{mpc} to determine the dominating \acp{mpc} reflecting off ambient scatterers (e.g., walls, doors, and the ceiling) and their corresponding \ac{aoa}. The next subsection presents our approach.

\subsection{Clustering of the multi-path components} \label{subsec:clustering}

Although \ac{mmw} signals propagate quasi-optically, diffuse paths can account for up to 40\% of the channel energy~\cite{9664350,8581503}. Diffuse paths tend to aggregate, 
forming dense clusters in both the angle and delay domains.
Hence, \ac{mmw} channel models cluster the \acp{mpc} over both domains~\cite{molisch2012wireless, 8319437}.

We employ clustering to group closely-packed \acp{mpc} that reach the receiver. We achieve this through \ac{dbscan}, a density-based algorithm that determines clusters by finding data points lying within a user-defined radius and sharing a given number of neighbors~\cite{ester1996density}. A data point includes the time delay, azimuthal \ac{aoa}, and elevation \ac{aoa} features of the \acp{mpc}~\cite{mpcTAP2018Camillo}. 
Unlike other conventional clustering approaches such as K-means and \ac{knn}, \ac{dbscan} does not require prior information on the number of clusters. It can identify clusters of different shape and size, and can separate dense \ac{mpc} clusters from sparsely spread-out \acp{mpc}.
Each cluster formed is likely reflected off different ambient scatterers. As a result, the number of clusters is also an estimate of the number of reflecting surfaces.
To manage the few cases where \ac{dbscan} clusters multiple \acp{mpc} together upon a single pass, we use \ac{dbscan} recursively. Namely, for each cluster, we re-run the algorithm with a progressively smaller cluster radius, and allowing a smaller number of neighbors in the cluster. 

We summarize the clustering procedure in Algorithm~\ref{alg:recDBSCAN}. Consider data from one \ac{tx} for now. 
Call ${\bf X}_t = [\pmb{\tau}, \pmb{\theta}, \pmb{\psi}]$ the matrix containing the features of all $L_t$ \acp{mpc} that the \ac{rx} observes at the $t$-th location, where the $L_t \times 1$ vectors $\pmb{\tau}, \pmb{\theta}, \pmb{\psi}$ represent the delay (in ns), azimuth, and elevation \acp{aoa} (in degrees wrapped in the interval $[0^{\circ}, 360^{\circ})$) of the \acp{mpc}, respectively. This yields a similar scale for all values, and does not require additional feature rescaling.

We initially cluster ${\bf X}_t$ with \ac{dbscan}, resulting in a set of clusters  $\mathcal{X}_{t}$ (line~\ref{ln:dbscan}). We compute the centroids using the core points of each cluster in $\mathcal{X}_{t}$ (i.e., those cluster points that share at least $\gamma$ neighbors) and store them in  set $\mathcal{C}$ (line~3). Then, for each cluster and its centroid, we compute a cost defined as the average Euclidean distance between the centroid and each of cluster's points. If this cost is greater than the cluster radius $\epsilon$ (line 4), we remove the centroid of the cluster from set $\mathcal{C}$ (line 7), and re-cluster the points through \ac{dbscan} (line~8), with a radius and number of neighbors scaled down by $\eta$ (line~5). We then add the centroids of the newly determined clusters back to set $\mathcal{C}$ (lines~10-11).

We repeat the above process until all clusters are composed of points located no farther from the centroid than the cluster radius parameter in the delay-\ac{aoa} domain. The resulting set of centroids $\mathcal{C}$ corresponds to the dominant \acp{mpc} (both \ac{los}, if available, and \ac{nlos}) reflected off surfaces or obstacles and observed at the \ac{rx} location. The presence of multiple points in each cluster represents dispersed/scattered components related to the same \ac{mpc}~\cite{clusteringVTC2017Jian}.
After clustering \acp{mpc} from all transmitters $t$, we group the resulting cluster sets $\mathcal{X}_t$. The centroids are finally characterized by their mean \ac{mpc} delay and mean azimuth and elevation \acp{aoa}.

\remembertext{dbscanComplexity}{We remark that the average complexity of a regular \ac{dbscan} procedure is $\mathcal{O}(N \log N)$, where $N$ is the total number of \acp{mpc} obtained from each \ac{tx} (in the order of few hundreds). On the contrary, the complexity of recursive \ac{dbscan} procedure is $\mathcal{O}(N \log N + K \cdot m \log m)$, where $K$ is the upper-bound average number of clusters to be re-clustered and $m$ is the average number of \acp{mpc} in such cluster. As our proposed procedure only breaks down those clusters that have a total associated cost exceeding $\epsilon$, $K$ and $m$ are significantly less than $N$. Thus, the increase in the average complexity remains relatively insignificant. }

\remembertext{dynamic}{Dynamic changes in the environment can affect small-scale fading parameters such as phase and \ac{mpc} delay~\cite{channelParams2022Access}. However, recursively clustering these \acp{mpc} and subsequently computing the centroids helps average-out these fading effects. As these centroids eventually correspond to the \acp{aoa} (and thus, the \acp{adoa}) of the dominant \acp{mpc}, our localization schemes can thus be resilient to dynamic environments.}




To compute \acp{adoa}, we first choose the reference \ac{mpc} $\bar{\bf c}_{k^\star}$ as the one with the least delay, where $k^\star = \argmin_k \tau_k$ for $k=1,2,\cdots,|\mathcal{C}|$. We then remove $\bar{\bf c}_{k^\star}$ from $\mathcal{C}$ and sort the \acp{aoa} of the remaining centroids in increasing order. In practice, this corresponds to sweeping the entire azimuthal angle domain in increasing order of the \acp{aoa} (as seen from the global reference frame). We then subtract $\theta_{k^\star}$ from the sorted \acp{aoa}. We finally form the input feature vector for the \ac{nn}, by expressing all \acp{adoa} in radians wrapped in $[-\pi,\pi)$.

\begin{algorithm}[t]
\renewcommand{\baselinestretch}{1.15}
\SetKwInput{KwInput}{Input}                
\SetKwInput{KwOutput}{Output}              
\DontPrintSemicolon

  
  
  \SetKwProg{Fn}{Function}{:}{}
  \Fn{\textsc{Rec\_DBSCAN}\,($\bf{X}$, $\epsilon$, $\gamma$, $\eta$)}{
        $\mathcal{X} \gets$ DBSCAN\,$({\bf X}, \epsilon, \gamma)$\; \label{ln:dbscan}
        $\mathcal{C} \gets$ \textsc{Centroid}\,$(\mathcal{X})$\; \label{ln:centroid}
        
        \While{$\exists \, {\bf X}_c\in \mathcal{X}$ of centroid ${\bf c}$ s.t.\ \textsc{Cost}\,$({\bf{X}}_c,{\bf c}) > \epsilon$}{
            $\epsilon' \gets \eta\epsilon$\, , \,\,\,  $\gamma' \gets \eta\gamma$\;
            \ForEach{${\bf X}_c\in \mathcal{X}$ of centroid ${\bf c}$ s.t.\ \textsc{Cost}\,$({\bf X}_c
            ,{\bf c}) > \epsilon'$ }{
                $\mathcal{C} \gets \mathcal{C} \smallsetminus \{{\bf c}\}$\;
                $\mathcal{X}_{n} \gets$ DBSCAN\,$({\bf X}_c, \epsilon', \gamma')$\;
                \ForEach{ ${\bf X}_n \in \mathcal{X}_{n}$ }{
                    ${\bf c}_n \gets$ \textsc{Centroid}\,$({\bf X}_n)$\;
                    $\mathcal{C} \gets$ Append$( {\bf c}_n )$\;
                }
            }
            $\epsilon \gets \epsilon' \,$ , \,\,\,  $\gamma \gets \gamma'$\;
        }
        \KwRet $\mathcal{C}$\;
  }
  \caption{Recursive DBSCAN}
  \label{alg:recDBSCAN}
  \end{algorithm}

\subsection{Architecture of the NN}
\label{sec:algo.nnmodel}

We propose tiny \acp{nn} with four layers. The number of neurons of the first layer matches the maximum number of \acp{aoa} we expect, i.e., $N_i = N_a-1$. The first hidden layer contains $N_{h_1} = \kappa N_i$ neurons. For the second hidden layer, $N_{h_2} = N_{h_1}$, whereas for the third hidden layer we set $N_{h_3}=N_{h_2}/2$. Finally, the output layer consists of two neurons, conveying the regression estimate of the 2D coordinates of the \ac{rx}. 
Table~\ref{nnModelparam} summarizes the number of neurons in each layer, where $\lceil{\cdot}\rceil$ represents the ceiling function.

\begin{table}[t]
    \renewcommand{\arraystretch}{1.15}
    \caption{Summary of the number of neurons in each layer}
    \label{nnModelparam}
    \centering \footnotesize
    \begin{tabular}{@{\hspace{1mm}}lc@{\hspace{1mm}}}
        \toprule
        \textbf{Layer} & \textbf{Number of neurons} \\
        \midrule
         Input layer~($N_i$) & $N_a - 1$  \\
         Hidden layer 1~($N_{h_1}$) & $\lceil{\kappa N_i \rceil}$ \\
         Hidden layer 2~($N_{h_2}$)  & $ N_{h_1} $\\
         Hidden layer 3~($N_{h_3}$)  & $\lceil N_{h_2}/2 \rceil $\\
         Output layer~($N_o$) & $2$ \\
        \bottomrule
    \end{tabular}
    \vspace{-2mm}
\end{table}

We train our \ac{nn} to learn a non-linear function that maps the input \acp{adoa} to the output \ac{rx} coordinates. 
Call ${\bf x}$,  the \ac{rx} location coordinates, ${\Theta}_{\bf x}$ the input \ac{adoa} vector at location ${\bf x}$, and ${\bf W}_i = [{\bf w}_1 \cdots {\bf w}_{n_i}]$ the weight matrix of layer $i$,
where $n_i$ represents the number of neurons in layer $i$, and each $n_{i-1} \times 1$ vector ${\bf w}_j$ contains the weights of the edges that connect the $n_{i-1}$ neurons in layer $i-1$ to the $j$th neuron in layer $i$. Furthermore, let ${\bf b}_i = [b_1 \cdots b_{n_i}]^T$ be the vector containing the bias values for each neuron in layer $i$, and call $\mathcal{A}(\cdot)$ the activation function for each neuron. Then, the output values of layer $i$ can be expressed recursively as
\begin{equation}
    \mathbf{y}_i = \mathcal{A}({\bf W}_i^T {\bf y}_{i-1} + \mathbf{b}_i) \; ,
\end{equation}
where applying $\mathcal{A}(\cdot)$ to a vector denotes computing the activation function $\mathcal{A}(\cdot)$ for each entry of the vector, and yields the vector of the corresponding results. Note that ${\bf y}_0 = {\Theta}_{\bf x}$ represents the input to the \ac{nn}.
As our \acp{nn} models have $n=4$ layers, the estimated location of the receiver is given by
\begin{equation}
    \hat{\bf x} = [\hat{x}_1,\hat{x}_2]^T = \mathbf{y}_4 = \mathcal{F}({\Theta}_{\bf x}) \; , 
\end{equation}
where
\begin{dmath}
    \mathcal{F}({\Theta}_{\bf x}) = \mathcal{A}\big({\bf W}_4^T\mathcal{A}\big( {\bf W}_3^T \mathcal{A}\big( {\bf W}_2^T \mathcal{A}({\bf W}_1^T{\Theta}_{\bf x}+{\bf b}_1)  + {\bf b}_2\big) +{\bf b}_3 \big) +{\bf b}_4 \big)
\end{dmath}
is the non-linear regression function the \ac{nn} applies to the \acp{adoa}. As the number of neurons in our \acp{nn} is small, computing ${\cal F}(\cdot)$ only requires a limited number of small-size matrix multiplications and vector summations. Such operations are feasible even for computationally-constrained devices. 
We choose the rectified linear unit activation function (ReLU) for $\mathcal{A}(\cdot)$, and train the network using the Adam optimizer.\glsunset{mse}
We frame the regression as a \acrfull{mse} minimization problem, and utilize the \ac{mse} loss function
\begin{equation}
    \mathcal{L}({\bf x},{\hat{\bf x}}) = |x_1 - \hat{x}_1|^2 + |x_2 - \hat{x}_2|^2 \; ,
\label{regressionLoss}   
\end{equation}
where $(x_1,x_2)$ are the algorithm-supervised location labels for \ac{adoa} features from the bootstrapping algorithm, and $(\hat{x}_1,\hat{x}_2)$ are the location estimates yielded by the \ac{nn}.

\subsection{Hyperparameter tuning}

\begin{table}[t]
    \renewcommand{\arraystretch}{1.15}
    \caption{Summary of the hyperparameters
    }
    \label{hyperparameters}
    \centering \footnotesize
    \begin{tabular}{@{\hspace{1mm}}lc@{\hspace{1mm}}}
        \toprule
        \textbf{Hyperparameter} & \textbf{Range} \\
        \midrule
        Node factor~($\kappa$) & $\{0.6,~0.7,~0.8,~0.9,~1\}$  \\
        Dropout rate ($p$) & $\{0\%,~10\%,~15\%,~20\%\}$ \\
        Learning rate ($r$) & $[0.0001, \ldots, 0.01]$\\
        Batch size~($b$) & \{50\%, 75\%\}\\
        \bottomrule
    \end{tabular}
    \vspace{-2mm}
\end{table}

In order to optimize the \ac{nn} to minimize the loss function, we automatically tune the following hyperparameters of the model: 
i) the node factor $\kappa$, which defines the number of neurons in the hidden layers; ii) the dropout rate $p$, which helps prevent overfitting by ignoring $p$\% of the \ac{nn} links during the training phase; iii) the learning rate $r$; 
and iv) the training data batch size $b$. 
\remembertext{hyperparameters}{Table~\ref{hyperparameters} summarizes the hyperparameter search ranges. The learning rate values range from $10^{-4}$ to $10^{-2}$ in a logarithmic progression, with ten values per decade.
The optimal hyperparameters for the \ac{nn} model are chosen from given set after an exhaustive grid search technique.}

\subsection{Choice of the bootstrapping localization algorithm}

\remembertext{input2}{In our algorithm-supervised approach, the bootstrapping algorithm has two objectives: ($i$) to localize a device while the \ac{nn} is not yet trained; ($ii$) to provide location labels that we can associate to \ac{adoa} measurements in order to train the \ac{nn}. Ideally, the algorithm should work with the same input data as the \ac{nn}, so that the devices does not need to measure/process additional metrics.}

In this work, we choose JADE~\cite{palacios2017jade}. Besides employing \acp{adoa} as input, JADE does not require knowledge of the surrounding environment or the location of the \acp{tx}, and is designed to jointly estimate both the location of the \ac{rx} and the location of all physical and virtual \acp{tx}. As the \ac{rx} moves, JADE progressively cumulates additional \ac{adoa} measurements, and becomes more accurate at estimating current and the past \ac{rx} locations, as well as anchor locations. The latter is also the main downside of JADE, as refining past estimates implies a quadratic increase in the complexity of the algorithm with the number of measurements. However, JADE still fits our case perfectly, since its location estimates enable us to train the \ac{nn} and switch to it well before JADE's complexity starts escalating~\cite{anish2022wcnc}.

\section{mmWave channel measurements from a 60 GHz sounder}
\label{sec:dataset}

To validate our localization scheme in a real indoor environment, we employ an experimental dataset collected at the NIST premises in Boulder, CO, USA. In this section, we describe the dataset collection procedures and the pre-processing steps required to extract \acp{adoa}.

Firstly, we outline the~\ac{mmw} channel sounder used to capture the channel measurements. We then describe the measurement campaign and the \ac{mpc} extraction technique. Finally, we discuss the pre-processing steps that yield the \acp{adoa} of the \acp{mpc}.

\begin{figure*}
    \centering
    \newlength{\channelexpheight}
    \setlength{\channelexpheight}{6.25cm}
    \subfloat[\label{fig:environment}]{\includegraphics[height=\channelexpheight-0.2cm]{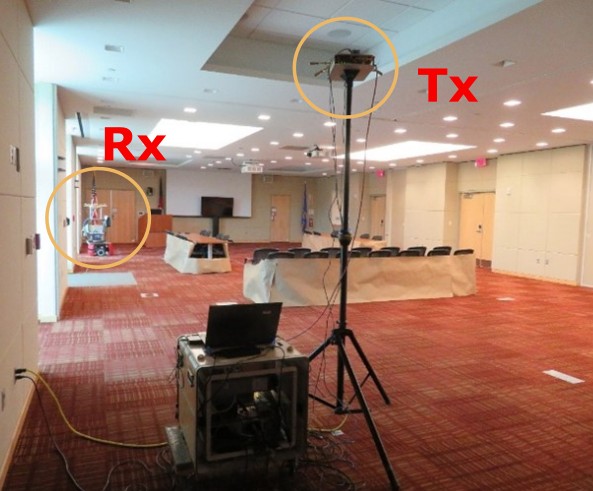}}
    \hfill
    \subfloat[\label{fig:map}]{\includegraphics[height=\channelexpheight-0.1cm]{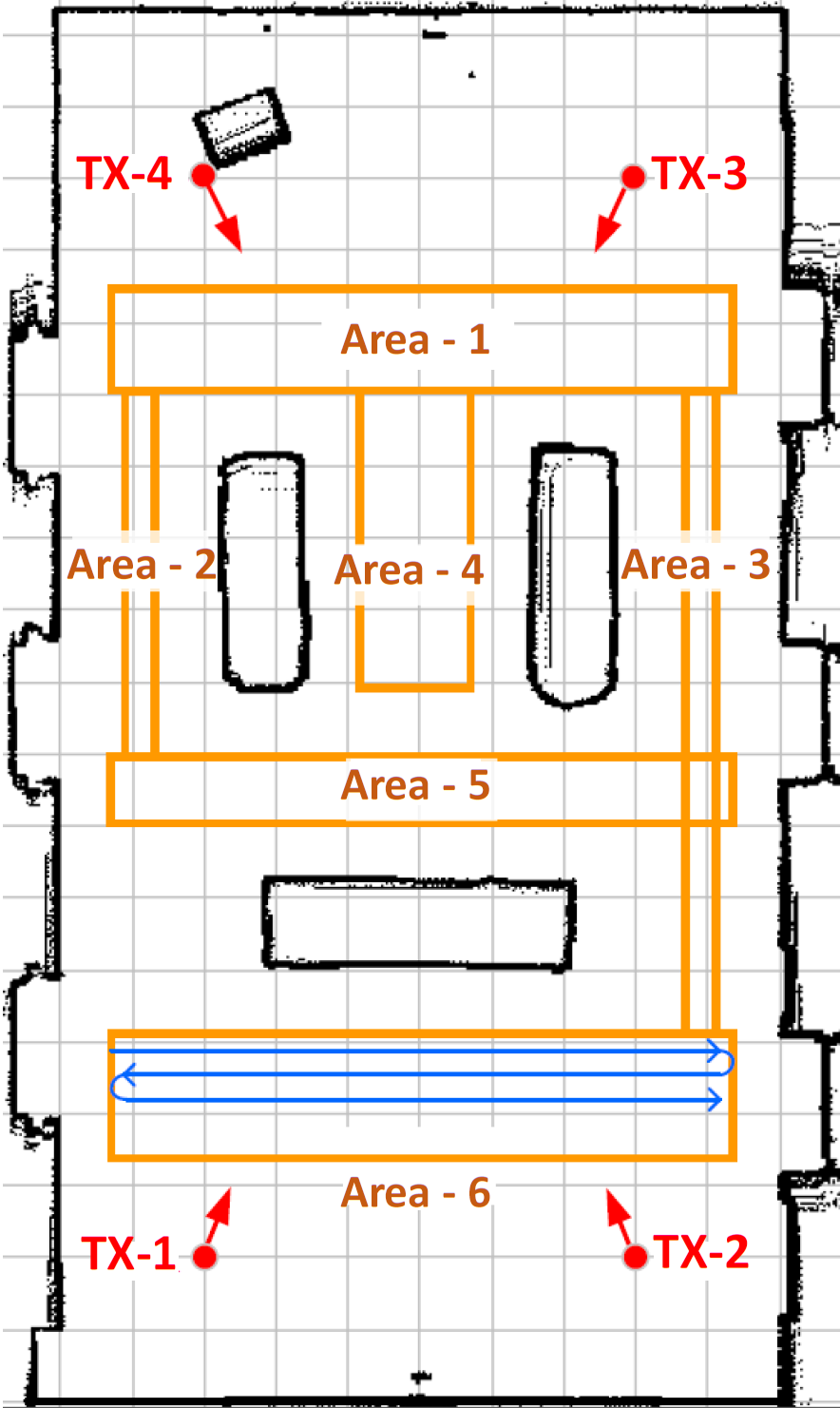}}
    \hfill
    \subfloat[\label{fig:mapfig}]{\includegraphics[height=\channelexpheight]{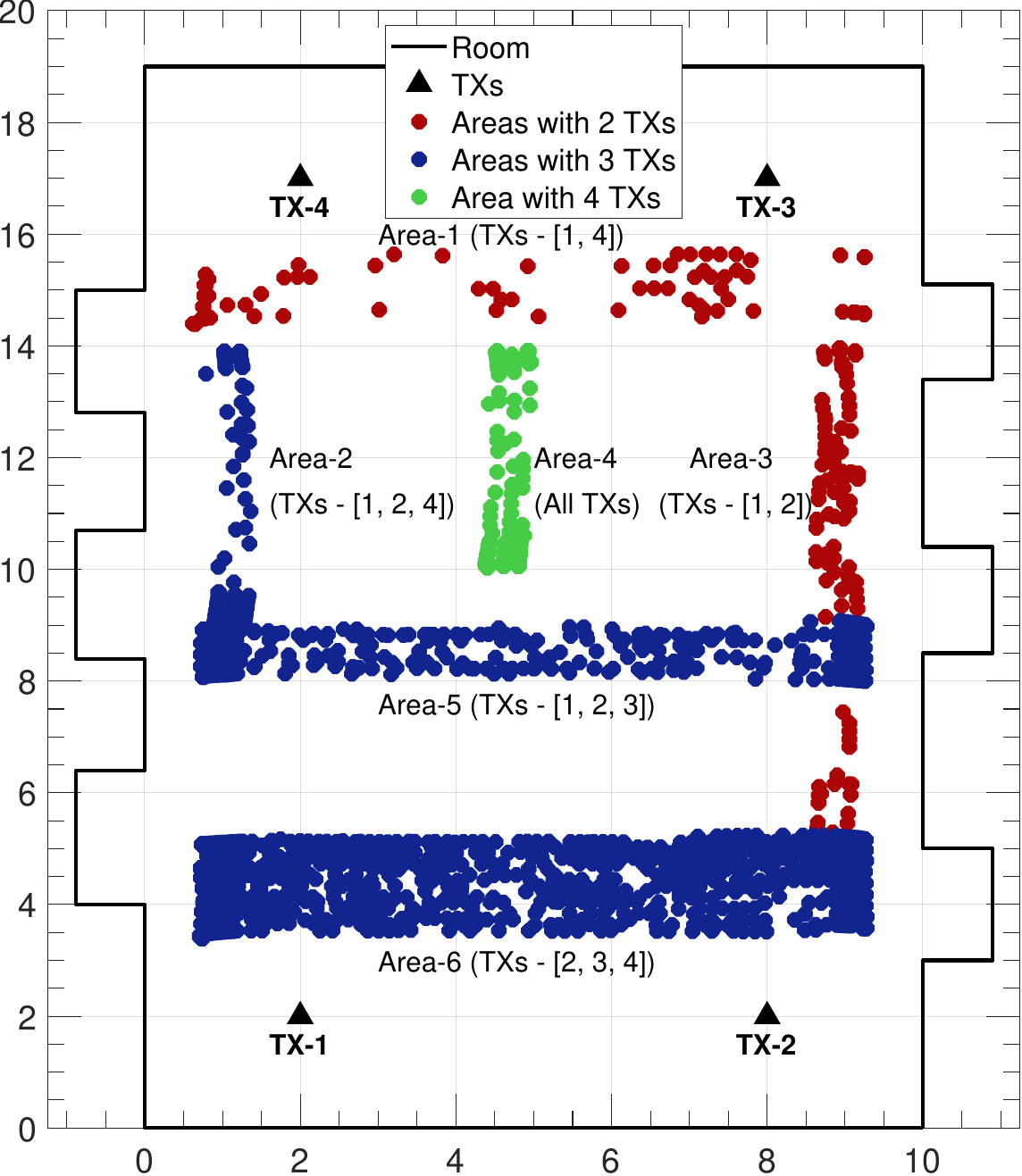}}
    \caption{(a) Measurement setup and (b) floor plan of the room for the 3D double directional 60\,GHz channel sounder experiment conducted at NIST to collect indoor mmWave channel responses. (c) Dataset resulting from processing the channel estimation data. The areas are divided and annotated based on the number of \acp{tx} from which the \ac{rx} captures the signal. 
    Axis unit: [m]. Note the different axis scale.}
\end{figure*}


\subsection{Channel sounder}
Fig.~\ref{fig:environment} displays the \ac{tx} and the \ac{rx} of the NIST 60\,GHz 3D double-directional switched-array channel sounder. The \ac{rx} embeds a circular array of 16~horn antennas having a 3D Gaussian radiation pattern with 22.5$^\circ$ beamwidth and 18.1\,dBi gain. To avoid ``blind spots'', the angular spacing between adjacent horns along the array's azimuthal plane is matched to the beamwidth, i.e., subsequent horns are oriented 22.5$^\circ$ away from each other. Moreover every second horn in the circular array is oriented towards an elevation angle of 22.5$^\circ$. The resulting synthesized azimuth field-of-view of the array is 360$^\circ$, whereas the elevation \ac{fov} is 45$^\circ$.

The \ac{tx} is almost identical, except that it embeds a semi-circular array with 8 horns, limiting the azimuthal \ac{fov} to 180$^\circ$ while maintaining the same 45$^\circ$-elevation \ac{fov}. We synchronously trigger the \ac{tx} and the \ac{rx} via rubidium clocks at each end, which are also used to discipline the local oscillators. For further system details, we refer the interested readers to~\cite{7928270,9664350}.

\subsection{Measurement campaign and MPC extraction}
\label{subsec:campaign}

The measurements  took place in a 19.3\,m~$\times$~10\,m lecture room (Fig.~\ref{fig:map}). The \ac{tx} was installed on a tripod at a height of 2.5\,m, whereas the \ac{rx} was mounted on a mobile robot at a height of 1.6\,m. An onboard computer on the robot makes it possible autonomously record channel acquisitions, and a laser-guided navigation system reports the robot's ground-truth location and heading. The top half of the room in Fig.~\ref{fig:environment} included two tables with surrounding chairs, while the bottom half included two rows of chairs facing the top wall. 

During the measurements, the \ac{tx} was iteratively placed at the four corners of the room, marked \mbox{TX-1} to TX-4, and oriented towards the center of the room (indicated by the arrows) in order to optimize the spatial coverage of the \ac{mmw} signal. 
For each \ac{tx} location, the \ac{rx} moved along a lawnmower trajectory in each of the measurement areas labeled 1 to 6 at a maximum speed of 2\,m/s. The average distance between consecutive recordings was $\sim$9\,cm, which emulates measurements taken during \ac{rx} motion. The channel was void of any motion, e.g., humans in the room, except for that of the \ac{rx}. We recorded 10\,895 channels in total.

Each acquisition sequentially measures the \ac{cir} between each of the 8~\ac{tx} and each of the 16~\ac{rx} antennas, resulting in 128 \acp{cir} per acquisition.
The 128 \acp{cir} are then coherently combined through the \ac{sage} super-resolution algorithm~\cite{hausmair2010sage} to extract \acp{mpc} channel paths and their geometric properties.
While the SAGE algorithm de-embedded the directional beam patterns of the antennas, we removed the effects of the \ac{tx} and \ac{rx}'s front ends via pre-distortion filters~\cite{mpcTAP2018Camillo,9664350}.

\subsection{Dataset preprocessing}
\label{subsec:dataset.preprocess}

We pre-process the collected dataset to compute \acp{adoa} that can input into our localization scheme.
The rationale behind this step is that, ideally, a \ac{rx} should be able to measure \acp{mpc} from all \acp{tx} at the same location. Instead, the measurements the \ac{rx} takes from different \acp{tx} are slightly displaced. This is because the robot that moves the receiver needs to be restored to its starting location and start its path anew every time the \ac{tx} is relocated. To overcome this shortcoming, we seek closest locations where the \ac{rx} obtained measurements from different transmitters, and coalesce them into a single set of \acp{mpc} measured at roughly the same location.
For this, we use the range searching technique as follows. 

Let $\mathcal{M}_{n,t}^i$ be the set of all locations in area $n$ where the \ac{rx} takes channel measurements from \ac{tx} $i \in \mathcal{T}_n$, where $\mathcal{T}_n$ is the set of transmitters available in area $n$. These locations are indexed by time step $t = 1,2,\cdots,N_n^i$, where $N_n^i$ is the total number of measurement locations for transmitter $i$ in area $n$.
Every location ${\bf m}_{n,t}^i \in \mathcal{M}_{n,t}^i$ is a 3D coordinate vector associated with a measurement matrix $\mathbf{\Gamma}_{n,t}^i = [\pmb{\alpha}, \pmb{\tau}, \pmb{\theta}, \pmb{\psi},\pmb{\phi}, \pmb{\beta}]$
of size $C_{n,t}^i \times 6$, where $C_{n,t}^i$ is the number of \acp{mpc} returned by the channel estimation algorithm. In $\mathbf{\Gamma}_{n,t}^i$, $\pmb{\alpha}$ contains the path loss of each \ac{mpc} (in dB), $\pmb{\tau}$ represents the propagation delay (in ns), $\pmb{\theta}$ and $\pmb{\psi}$ are the azimuth and elevation \acp{aoa} at the \ac{rx} (in degrees), whereas $\pmb{\phi}$ and $\pmb{\beta}$ are the azimuth and elevation \acp{aod} at the \ac{tx} (in degrees).

To carry out the range search, we first elect a reference \ac{tx}, e.g., the one with the lowest ID number. Call this \ac{tx} $i^\star$. Then we start from $t=1$ and find the closest measurement locations for all remaining available transmitters. Namely, $\forall i \in \mathcal{T}_n \smallsetminus \{i^\star\}$, we find
\begin{equation}
    u_i^\star = \argmin_{u_i} \big\| {\bf m}_{n,u_i}^i - {\bf m}_{n,t}^{i^{\star}} \big\| \, .
\end{equation}
We then cluster the measurement locations as set
\begin{equation}
    \mathcal{C}_{n,t} = \big\{ {\bf m}_{n,t}^{i^{\star}} \big\} \cup \Big\{ {\bf m}_{n,u_i^\star}^i\, , \, i \in \mathcal{T}_n \smallsetminus \{i^\star\} \Big\} \, .
\end{equation}
Finally, we remove the locations in $\mathcal{C}_{n,t}$ 
to avoid that they become also part of other sets formed later on, i.e.,
\begin{equation}
    \mathcal{M}_{n,t+1}^i = \mathcal{M}_{n,t}^i \smallsetminus \big\{ {\bf m}_{n,u_i^\star}^i \big\} \, .
\end{equation}
We repeat the above operations for all $t$, and finally take the centroid of each cluster ${\bf c}_{n,t}$ as the closest approximation of a measurement location having data from all \acp{tx} in area $n$, and merge all \acp{mpc} from all corresponding measurements in matrix $\mathbf{\Gamma}_{n,t} = [ \mathbf{\Gamma}_{n,t}^{'\,i^\star} \mathbf{\Gamma}_{n,t}^{'\,i^\star+1} \cdots \mathbf{\Gamma}_{n,t}^{'\, |\mathcal{T}_n|}]'$. 
Note that the number of \acp{mpc} will differ for each \ac{rx} location and for each \ac{tx}.


The above operation results in extracting 3\,595 \ac{rx} locations, as shown in Fig.~\ref{fig:mapfig} with one dot per location. The color codes refer to the number of transmitters from which the \ac{rx} records channel measurements (annotated in the plot). Note that the \ac{rx} records measurements from all \acp{tx} only when moving in area~4, whereas illumination from the four \acp{tx} is not uniform across the six areas. 



\begin{figure*}[t]
    \centering
    \newlength{\clustpanelheight}
    \setlength{\clustpanelheight}{5.25cm}
    \subfloat[\label{fig:rayTrace}]{\includegraphics[height=\clustpanelheight]{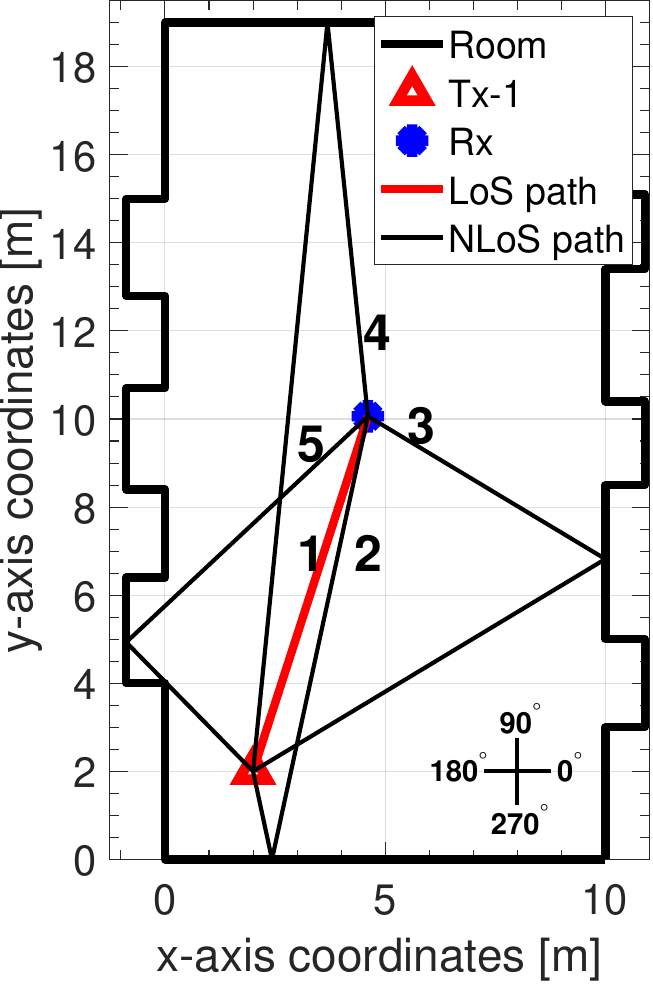}}\hspace{1cm} 
    \subfloat[\label{fig:cluster1}]{\includegraphics[height=\clustpanelheight]{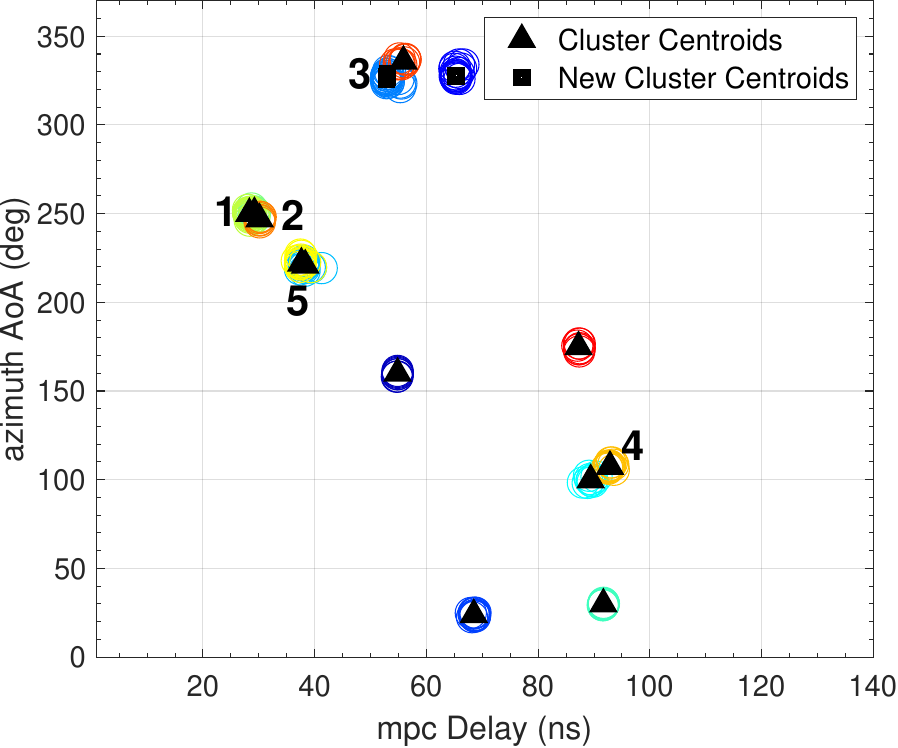}}
    %
    \caption{\acp{mpc} clustering using our proposed recursive approach for a sample \ac{rx} location in area 4. Note that the triangles represent the cluster centroids from the initial \ac{dbscan} procedure and the squares represent the centroids obtained after recursive clustering. (a) Reference ray traces of the LoS and first-order \acp{mpc} from \ac{tx}-1 to the \ac{rx}. 
    (b) Clustering of the \acp{mpc} from \ac{tx}-1.}
\end{figure*}


\section{Evaluation}  \label{sec:results}

We proceed to evaluate the performance of our proposed scheme. First, we use the data collected at the NIST Boulder's campus, as presented in the previous section. Then, we extend the analysis to other scenarios by means of simulation, in order to explore the distribution of localization errors across an indoor spacethe impact of the number of \acp{tx}, the standard deviation of the \ac{aoa} errors, and the size of the training dataset. 
We start the bootstrapping process by running the JADE algorithm on the full dataset. Due to uneven \ac{tx} illumination, we cannot collect a sufficient number of \acp{aoa} (and therefore \acp{adoa}) at some \ac{rx} locations, especially in area 1 and area 3. In these cases, JADE cannot localize the \acp{rx} due to insufficient data. The remaining points where JADE can find a fix are $\sim$2\,600.

\subsection{Illustration of the MPCs clustering operation}

We start by discussing the performance and results of the clustering process outlined in Section~\ref{subsec:clustering} and 
Algorithm~\ref{alg:recDBSCAN}. 
As an example, consider one \ac{rx} location within area~4. 
Fig.~\ref{fig:rayTrace} shows a ray trace from TX-1, where we show the \ac{los} \ac{mpc} and first-order reflections reaching the \ac{rx}. As expected, these \acp{mpc} spread over the full azimuthal angle domain due to the reflections from the perimeter of the room.

We compare this ray trace with the data from the measurements in Fig.~\ref{fig:cluster1}. Here, we illustrate the \acp{mpc} clustered from \ac{tx}-1 in the delay-angle domain for the same \ac{rx} location. Throughout our evaluation, we set the cluster radius $\epsilon = 3$, the number of neighbors $\gamma = 6$ and the scaling factor $\eta=0.75$, chosen after an exhaustive search. 
For clarity, we portray the \acp{mpc} only in the delay and azimuth \ac{aoa} domains, although we recall that elevation \ac{aoa} values are also used as clustering features.

In Fig.~\ref{fig:cluster1}, we mark cluster centroids as black triangles (after the initial \ac{dbscan} pass) and squares (after the recursive clustering procedure). As a first result, we observe that hierarchical reclustering is needed otherwise, e.g., the earliest arrivals having an \ac{aoa} of $300^\circ$ and above (in dark and light blue towards the top of the picture) would be mistaken as a single arrival. 
Each centroid can be interpreted as a dominant \ac{mpc}. We numbered the most significant ones to illustrate that identified clustering include all key \acp{mpc} expected from the ray trace in panel~\subref{fig:rayTrace}. The centroid corresponding to the smallest \ac{mpc} delay represents the \ac{los} path (ray 1 in Fig.~\ref{fig:rayTrace}), as confirmed by its \ac{aoa} of about $250^\circ$. Similarly, \ac{mpc}~2 has slightly larger \ac{aoa} and delay features with respect to \ac{mpc}~1; conversely, \ac{mpc}~4 has a comparatively longer path length and an \ac{aoa} of about $100^{\circ}$. The rest of the \acp{mpc} from TX-1 are annotated based on the scheme proposed in Section~\ref{subsec:clustering}. 

From the azimuthal \ac{aoa} features of the cluster centroids, we form the set of \acp{adoa} values for the localization algorithm.


\subsection{Performance evaluation for entire dataset}
\label{sec:results.entire}

We now evaluate the performance of our tiny \ac{nn}. We evaluate our model against JADE and other geometry-based localization schemes. 
For this, we exploit the entire set of receiver locations across the room. Specifically, we initially run JADE on the entire dataset of $\sim$2\,600 measurements to obtain the corresponding \ac{rx} location estimates. 
Even when working on datasets with many measurements, some of JADE's location estimates can be largely off. To avoid that such errors negatively affect the \ac{nn} training process, we compute the centroid of all location estimates, fit a Gaussian distribution to the location estimates, and remove the 5\% of all estimates that are farthest from the centroid. 
We then form a test set with 434 location labels and their corresponding \ac{adoa} features, and train the \ac{nn} over the remaining points. 

\begin{figure}[t]
    \centering
    \includegraphics[width=1\columnwidth]{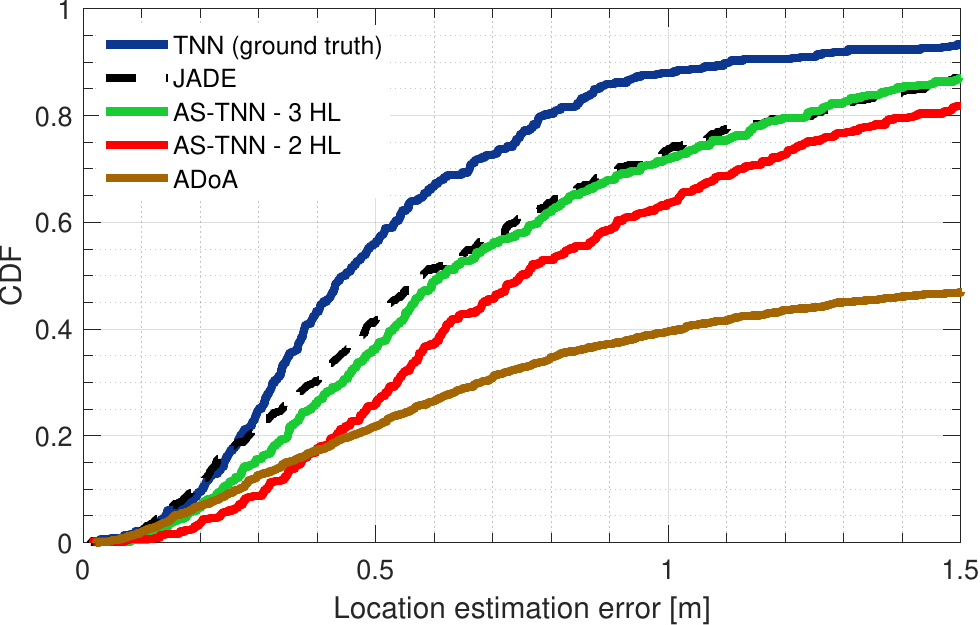}
    \caption{CDF of the location estimation errors for our proposed 3-hidden layer (HL) NN (referred to as AS-TNN) against the previous 2-hidden layer model and other state-of-the-art algorithms. Note that our tiny NN trained with ground truth is referred to as TNN. }
    \label{fig:cdfAll}
\end{figure}

\remembertext{nnGTvsJADE}{We illustrate the performance of our tiny \ac{nn} through the \ac{cdf} of the \ac{rx} location error in Fig.~\ref{fig:cdfAll}, where we also compare our algorithm-supervised approach (referred in the figures as AS-TNN) with the case when the \ac{nn} model is trained with ground-truth location labels (referred as TNN). Moreover, we also compare against JADE, the 2-hidden layer \ac{nn} model proposed in~\cite{anish2022wcnc}, and the geometric scheme ADOA~\cite{palacios2019single}. Our proposed \ac{nn} model has $(40,\ 36,\ 36,\ 18,\ 2)$ neurons respectively in each layer with the optimal hyperparameters $\kappa = 0.9$, $p = 0.2$, $r = 0.004$, and $b = 75\%$.  
We observe that our tiny algorithm-supervised \ac{nn} performs similar to JADE. Both schemes achieve sub-meter localization accuracy in about $74\%$ of the cases, with mean and median localization errors of about 86~cm and $\sim$61~cm, respectively. On the other hand, the previous 2-hidden layer version of the tiny \ac{nn} achieves sub-meter errors in 65\% of the estimates, with the corresponding mean and the median errors being 98~cm and $\sim$74~cm.

Our test environment entails a non-uniform distribution of \ac{rx} locations across the room and a different number of \ac{adoa} measurements in each room area. In this case, adding a hidden layer improves the localization accuracy by learning a more complex relationship between \ac{adoa} values and the location of the \ac{rx}. This contrasts the 2-layer \ac{nn} in~\cite{anish2022wcnc}, which worked well only with uniform datasets and regular room shapes. 
We observe that our proposed models as well as the zero-knowledge algorithm JADE outperform the geometric algorithm ADoA, that achieves sub-meter errors only in 40\% of the cases.

When trained with true \ac{rx} labels, the \ac{nn} outperforms JADE, with sub-meter accuracy in about $88\%$ of the cases, and a lower mean and median errors of 60~cm and $\sim$45~cm. We emphasize that training the \ac{nn} with ground-truth location labels is used here only to verify that the \ac{nn} learns to accurately map \ac{adoa} measurements to the corresponding \ac{rx} location. In practice, ground-truth locations would require a burdensome training dataset collection process, making our algorithm-supervised approach much more convenient.
While JADE and our \ac{nn} perform comparably in Fig.~\ref{fig:cdfAll}, JADE's complexity increases as new measurements flow in~\cite{palacios2017jade}, whereas the \ac{nn} only requires a limited set of matrix multiplications and vector additions computed on a small-sized input set, once the training is complete.}


\begin{figure*}[t]
    \centering
    \newlength{\coloreddotheight}
    \setlength{\coloreddotheight}{6cm}
    \subfloat[\label{fig:errorMap}]{\includegraphics[height=\coloreddotheight]{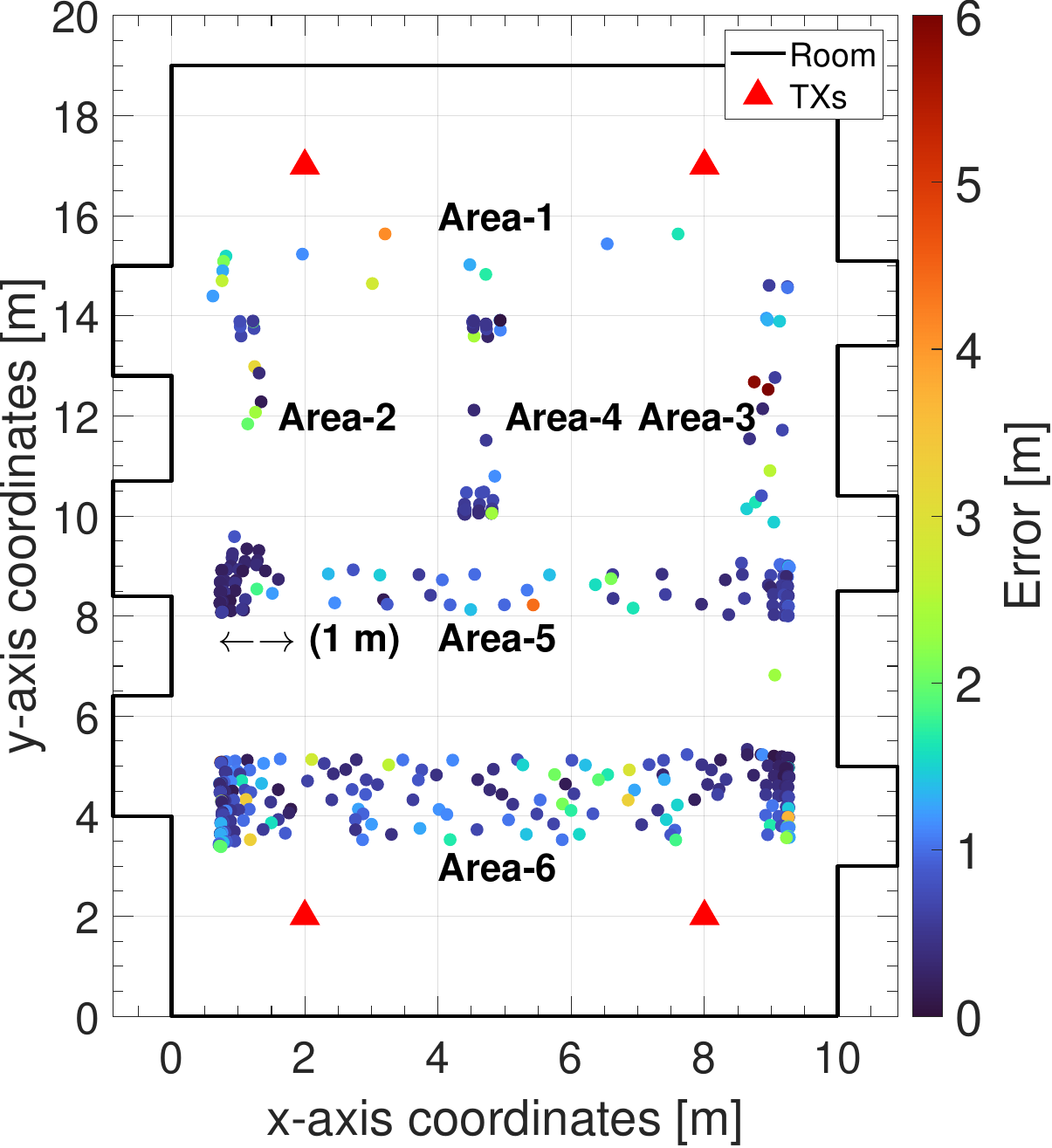}}
    \hspace{5mm}
    \subfloat[\label{fig:boxplotErr}]{\includegraphics[height=\coloreddotheight]{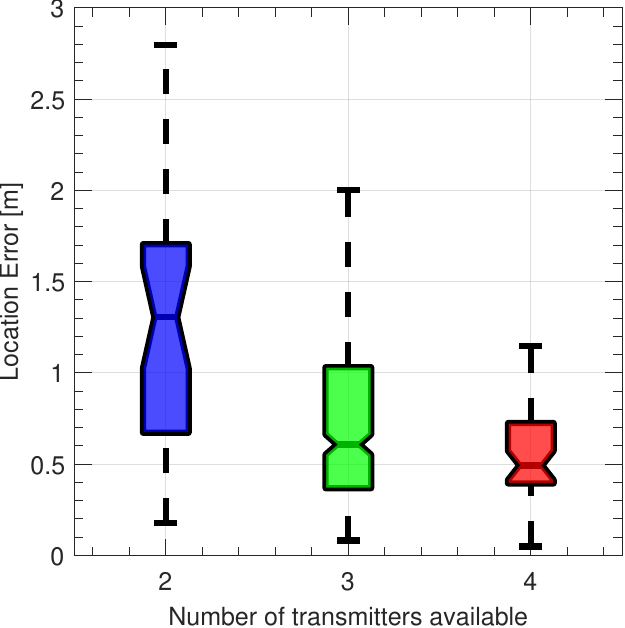}}
    \caption{Localization error for all regions with different number of transmitters available. (a) Location-wise performance of algorithm-supervised tiny \acp{nn} at the test locations (b) Statistical dispersion of the location error for regions having different numbers of available transmitters.}
\end{figure*}

To better understand the performance of our \ac{nn}, we examine how localization errors distribute over different locations across the room. Fig.~\ref{fig:errorMap} codes localization error at each test location via a colored dot, where blue hues represent low or acceptable errors. 
We observe localization errors of \textless{}1~m at the majority of the test locations, mostly in areas~2, 4, 5, and~6. There are two reasons behind this result. First, the \ac{rx} moving in these areas receives signals from 3 or 4 \acp{tx}. This yields a larger number of clusterized \acp{mpc}, thus a larger number of input \ac{adoa} values. The greater the number of \acp{adoa} per location, the more accurately JADE and the \ac{nn} can localize the \ac{rx}. Secondly, these areas include a larger number of \ac{rx} measurements available compared to areas 1 and 3, and constitute a larger proportion of the training data. Thus, the model can estimate the location of the \ac{rx} with better accuracy. On the contrary, large estimation errors in areas~1 and~3 tend to be larger. This is because the \ac{rx} records measurements only from two transmitters, and the set of measurement locations is sparser compared to other areas. However, many locations in these areas still yield sub-meter errors, implying that both JADE and our algorithm-supervised \ac{nn} infer the relationship between \acp{adoa} and \ac{rx} locations even by observing a limited number of examples. 

Fig.~\ref{fig:boxplotErr} summarizes the above observations by showing the statistical dispersion of the location estimation error for different regions, as a function of the number of available \acp{tx}. The median localization error decreases with increasing number of transmitters, and the greatest improvement is attained when transitioning from two to three available \acp{tx}. The dispersion of the error also decreases: the whiskers of the boxplots span progressively smaller intervals for increasing number of transmitters, and the highest errors occur only in areas with two transmitters.
Although the number of \ac{rx} locations in the training and test sets is heavily biased towards regions with three available \acp{tx}, we still observe a general trend of decreasing localization errors with increasing number of transmitters. This is because the bootstrapping algorithm JADE can localize the \ac{rx} more reliably when observing more \ac{adoa} measurements. In these cases, the \ac{nn} will also be able to map the \acp{adoa} to the \ac{rx} location more accurately.

\subsection{Performance evaluation over different areas}

We now discuss the performance of our \ac{nn} when trained with a smaller training dataset. To do this, we leverage the diversity of measurements recorded during the campaign, where both the number of available \acp{tx} and the number of measurements collected vary across different areas.
We choose three diverse sections: areas~2, 3 and 4, which occupy the top half of the room; area~5, that spans the room horizontally towards the middle; and area~6, which spans the bottom side. 
In areas~2, 3, and 4, the dataset includes measurements from 455 \ac{rx} locations. Note that we consider these areas together, as otherwise JADE would not observe sufficiently many measurements to yield accurate estimates. 
The number of measurements in area~5 is slightly higher, with 593 valid locations. Finally, 1\,396 \ac{rx} measurements are available in area~6.
In all the three scenarios, we enact a 20\%--80\% split between the test set and the training set.

\begin{figure}[t]
    \centering
    \includegraphics[width=\columnwidth]{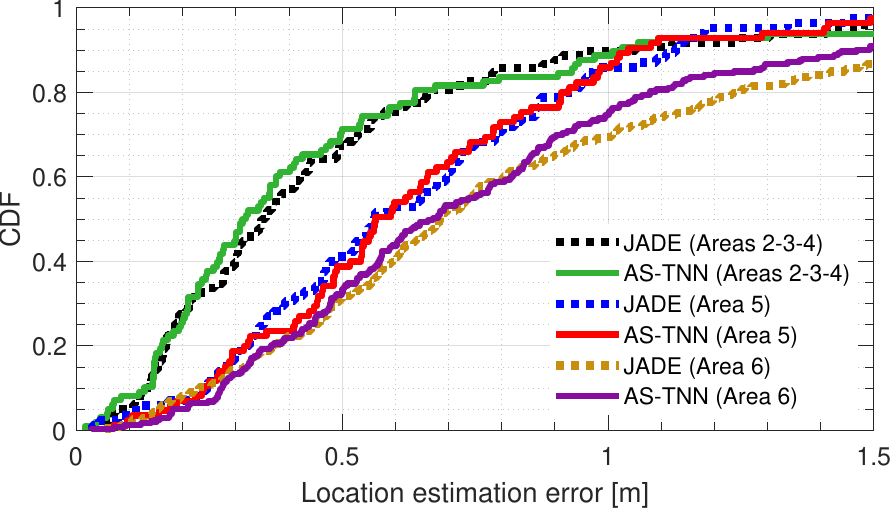}
    \caption{CDF of the location estimation errors for our proposed 4-layer NN when trained and tested in three scenarios: (i) in Areas 2, 3, and 4; (ii) in Area 5; and (iii) in Area 6.}
    \label{fig:cdfreg7345}
\end{figure}

Fig.~\ref{fig:cdfreg7345} shows the \ac{cdf} of the localization error attained by our \ac{nn} and by JADE. As previously observed, the \ac{nn} model learns the mapping between \acp{adoa} and the bootstrap location labels even with limited measurements, and matches the performance of JADE well in all scenarios. 
This opens the opportunity to train different tiny \acp{nn} for different areas, depending on the number of available \acp{tx}. This would improving the localization coverage, especially in large and distributed indoor environments. We leave this extension as a future development of our research.

Notice that the performance of both JADE and the \ac{nn} are better in areas~2, 3 and 4 as well as area~5, 
and more erroneous in area~6. 
This is a consequence of resorting to an algorithm-supervised approach, which inherently achieves the best performance only when the bootstrapping algorithm is also accurate throughout a given area. In our case, JADE performed remarkably good in areas~2, 3 and 4 as well as area~5. Here, the \ac{nn} is trained with low-error location labels, and therefore performs well. Conversely, JADE's location labels are affected by larger errors in area~6. This affects the training of the \ac{nn}, which achieves sub-meter errors in 75\% of the cases, instead of, e.g., 85\% for area~5. 

Compare now against the performance of the \ac{nn} trained with ground-truth location labels: we observe that the overall accuracy loss when the bootstrapping algorithm is accurate (such as in areas~2, 3 and 4, or in area~5) brings the proportion of locations with sub-meter errors from about $\approx$90\% to $\approx$85\%. If the bootstrapping algorithm is less accurate (e.g., like JADE working on data from area~6), the proportion decreases to about 75\%. 
Even thxough area~6 provides the largest number of dataset points, compounding training data from all areas increases the localization performance of the \ac{nn} with respect to training on data from area~6 alone.

\subsection{Impact of number of hidden layers and clustering of MPCs}
\label{sec:hiddenLay}
\begin{figure}[t]
    \centering
    \includegraphics[width=1\columnwidth]{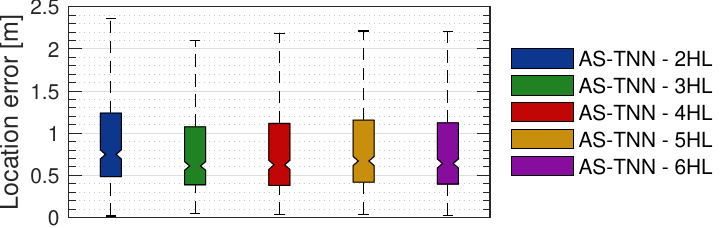}
    \caption{Statistical dispersion of localization errors for varying number of hidden layers in the tiny \ac{nn}. Here, AS-TNN - \emph{k}HL denotes our algorithm-supervised tiny NN with \emph{k} hidden layers.}
    \label{fig:boxplotTNN}
\end{figure}

\remembertext{boxTNN}{We now turn to assessing the impact of the number of hidden layers in the \ac{nn}, as this parameter affects the computational complexity. Fig.~\ref{fig:boxplotTNN} presents the statistical dispersion of localization errors for varying number of hidden layers in the tiny \ac{nn} model. Specifically, we vary from 2 hidden layers (as proposed in~\cite{anish2022wcnc}) to 6. We observe a decrease in the median error from $\sim$76~cm to $\sim$60~cm as we move from a 2-HL model to a 3-HL model. As more layers are added to the model, the whiskers of the boxplot span almost the same interval, implying a similar performance as a 3-HL model. We also note that, as the number of layers increases, the \ac{nn} will be able to represent a more complex regression function, but at a cost of learning and tuning more parameters. 
Moreover, adding hidden layers may lead to more accurate localization, it would also increase the computational complexity of \ac{nn} training and inference. As our primary goal is to learn the smallest \ac{nn} model that can match the performance of JADE, we limit our search to a 3-layer model.}

\begin{figure}[t]
    \centering
    \includegraphics[width=1\columnwidth]{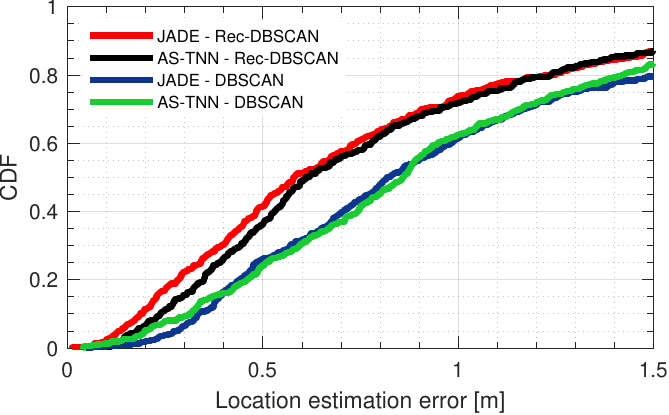}
    \caption{CDF of the location errors when using the dominant \acp{mpc} obtained from the regular \ac{dbscan} and the proposed recursive \ac{dbscan} (here, Rec-DBSCAN).}
    \label{fig:cdfDBSCAN}
\end{figure}

\remembertext{dbscan}{To understand the impact of clustering the \acp{mpc} on the localization performance, we propose to feed JADE and our \ac{nn} model with \acp{adoa} obtained after both regular and our proposed recursive \ac{dbscan} procedure. Fig.~\ref{fig:cdfDBSCAN} presents the \ac{cdf} of the localization errors incurred by JADE as well as our algorithm-supervised tiny \ac{nn} model when using \acp{adoa} obtained from the two procedures. While the \ac{nn} model is able to match the performance of the bootstrapping localization algorithm in both the cases, we can clearly observe an increase in the sub-meter localization errors from 61\% to 74\% when recursively clustering the \acp{mpc}. This is because recursive \ac{dbscan} finely breaks down large clusters, resulting in rich set of \ac{adoa} measurements corresponding to the dominant \acp{mpc} reflected off the  walls. This significantly improves JADE's performance, and subsequently that of the TNN.}

\subsection{Complexity analysis of the tiny NN}
\label{sec:complexity}

\remembertext{complexity}{%
We now evaluate the complexity of our \ac{nn} model using the following two metrics:

\textbf{Number of Parameters.} The computational complexity of a fully connected \ac{nn} is determined by the total number of parameters (i.e., the weights and biases) that the model learns during the training phase. For a fully connected \ac{nn}, the overall computational complexity is usually given as $\mathcal{O}(\sum_{k = 1}^{L} n_k \cdot m_k)$, where $n_k$ and $m_k$ are the number of input and output neurons in layer $k$ of the \ac{nn}. Our tiny \ac{nn} architecture comprises $(40,\ 36,\ 36,\ 18,\ 2)$ neurons respectively in each layer, resulting in an overall complexity of $\mathcal{O}(3420)$, and the number of trainable parameters equalling 3512. In case of our bootstrapping algorithm JADE, the complexity of the initial grid search problem is $\mathcal{O}(|\mathcal{A}|^2\, 2^{16})$, where $\mathcal{A}$ represents the total number of anchors. Each of the subsequent iterative MMSE problems have the quadratic complexity that depends on the number of anchors, number of iterations needed to converge, and the number of measurements. 

\textbf{Mean training and testing times.} We perform preliminary experiments to evaluate the mean training and inference times of our \ac{nn} model. For this, we used a Thinkpad E14 laptop, with an AMD Ryzen~7 4000 CPU and 24~GBytes of RAM, running in battery-saver mode. Our \ac{nn} model takes an average of $\sim$165~s to train, and the mean inference time is about 1.22~ms. Running JADE on the same machine takes about 1\,400~s (about 23 mins) to obtain the training labels. This is because JADE solves a large grid search problem having $2^{16}$ points, followed by computationally intensive iterative solution of two inter-dependent MMSE problems. Thus, from a systems perspective, we propose to offload JADE and the \ac{nn} training process to an edge server. 
}

\remembertext{complex_2}{
We remark that as the indoor scenario becomes more complex (i.e., with more obstacles, complex-shaped rooms, additional \acp{ap}, etc.), the number of \acp{mpc} between the \ac{tx} and the \ac{rx} will increase. To accurately localize the client, our bootstrapping localization algorithm will have to incorporate more measurements for each location, and process them altogether with a higher degree of computational complexity. As the input to both schemes is the same, the \ac{nn} model would require more parameters as well, in order to learn the mapping between the measurements and the location labels. This can be achieved by increasing the model's complexity, such as by adding more hidden layers or using convolutional layers. Note that, this will directly impact the computational complexity of the \acp{nn}. 
}

\subsection{Simulation study of localization error distribution}
\label{sec:results.sim}

\begin{figure*}[t]
    \centering
    \subfloat[\label{fig:Hroom}]{\includegraphics[height=4.6cm]{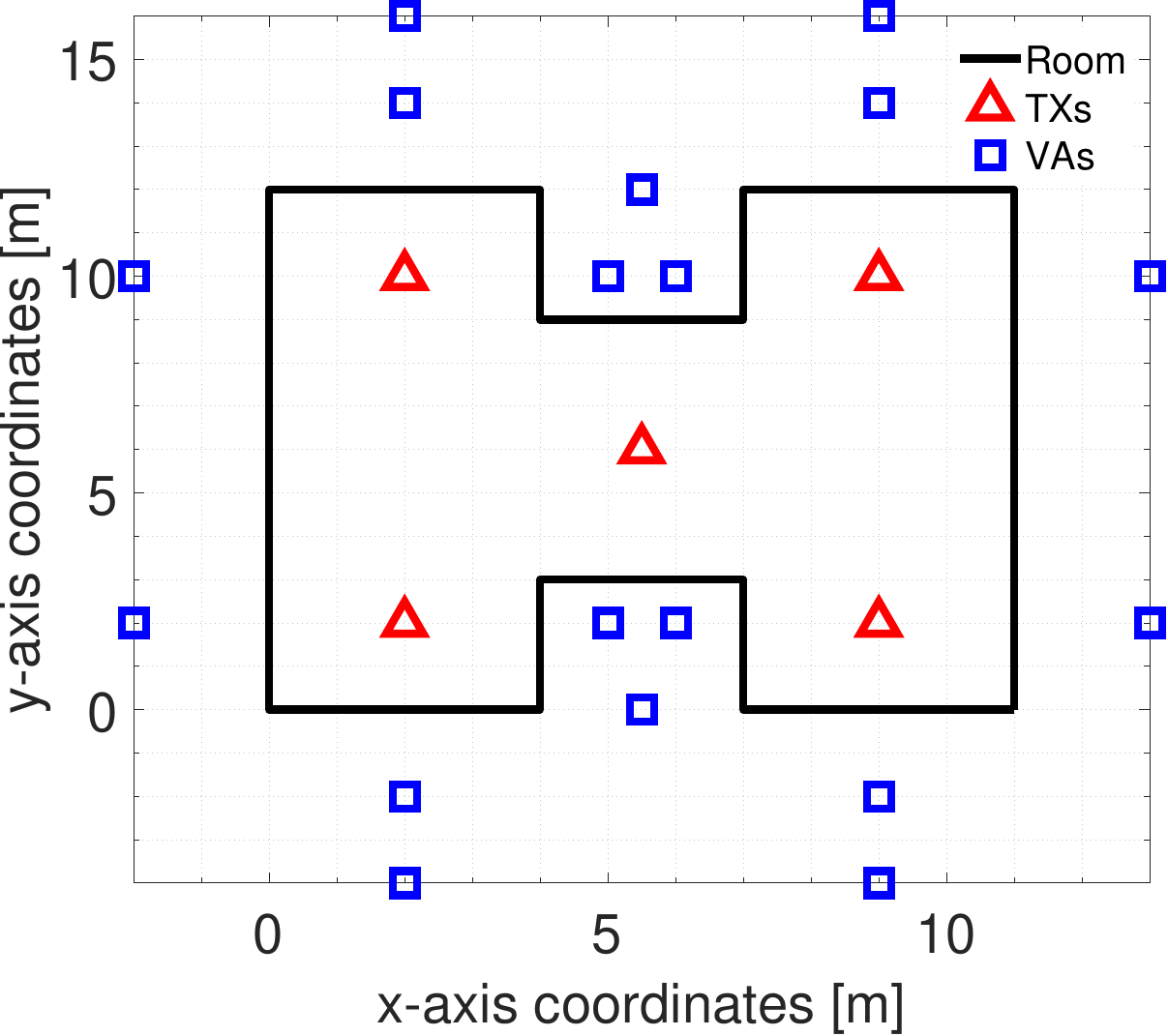}}
    \hfill
    \subfloat[\label{fig:cdfHsimData}]{\includegraphics[height=4.6cm]{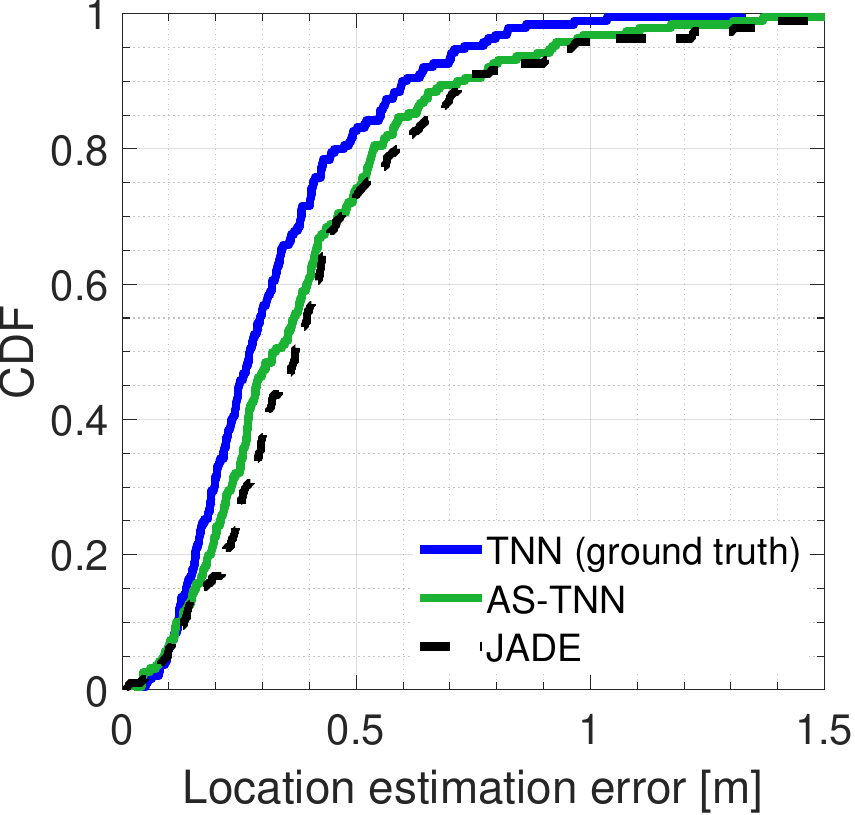}}
    \hfill
    \subfloat[\label{fig:HroomErrorMap}]{\includegraphics[height=4.7cm]{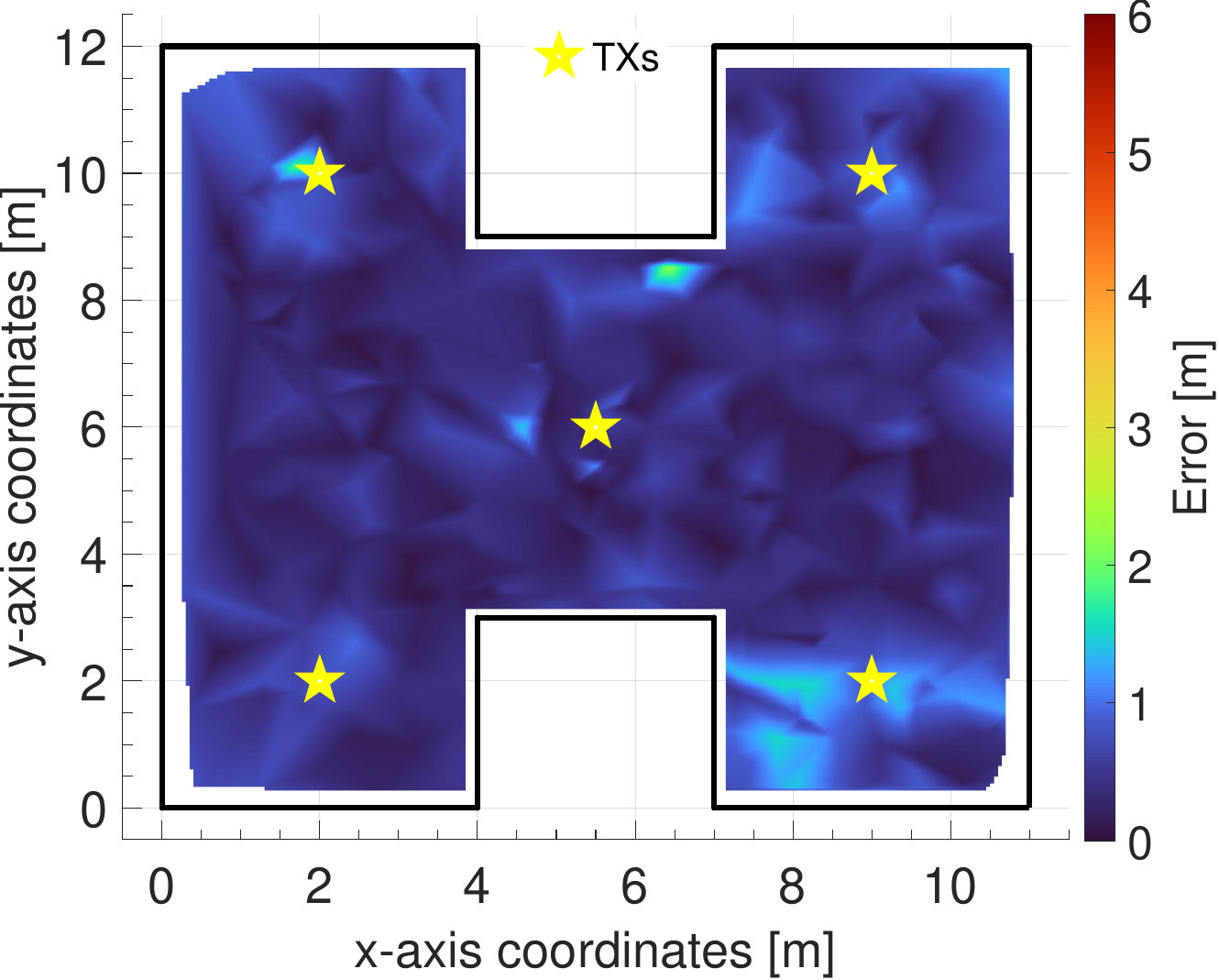}}

    \caption{(a) H-shaped room environment. Note the different axis scale. (b) CDF of location estimation error in the H-shaped room. (c) Heat map of the localization errors for the H-shaped room with 5 TXs. }
    \vspace{-4mm}
\end{figure*}

We now complement the above results by analyzing the performance of our model in an indoor environment via simulation. Fig.~\ref{fig:Hroom} illustrates an 11~m $\times$ 12~m H-shaped room consisting of two 4~m $\times$ 12~m rectangular sections and a 3~m $\times$ 6~m section in the middle. We deploy five \ac{mmw} \acp{tx} at coordinates $(2,2)$, $(2,10)$, $(5.5,6)$, $(9,2)$, and $(9,10)$. Blue squares convey a few first-order \acp{va} with respect to each wall of the room. 

We simulate a set of mobile \ac{rx} trajectories in the indoor environment to generate training data for the \ac{nn}. Each of the trajectories comprises 30 \ac{rx} locations. We collect the set of \acp{aoa} at each location from all transmitters and the \acp{va} using a ray tracer. In order to produce results that mimic realistic scenarios with imperfect \ac{aoa} estimation, we perturb the \acp{aoa} with zero-mean Gaussian noise of standard deviation $\sigma = 5^{\circ}$. 
In total, we generate 1\,500 \ac{rx} locations. A subset of these locations will be used to train and test our model as detailed below. 
%
%

%
As in Section~\ref{sec:results.entire}, we run JADE on the whole dataset and remove the 5\% of the location estimates that are farthest from the centroid of all estimates. This results in keeping~1\,425 client locations, of which 475 locations constitute the test set.
After hyperparameter optimization, the \ac{nn} model has $(40, 36, 36, 18, 2)$ neurons in each layer with $\kappa = 0.9$, $p = 0.1$, $r = 0.003$, and $b = 50\%$.

Fig.~\ref{fig:cdfHsimData} shows the \ac{cdf} of the location error as estimated using JADE and the algorithm-supervised \ac{nn}. We observe that the \ac{nn} performs as good as JADE with sub-meter location accuracy in about 92\% of the cases, with slightly better performance for errors less than $50$~cm. For comparison, 98\% of the estimates yield an error of \textless{}1~m when the \ac{nn} is trained with true locations. These results are significantly good for two main reasons: because JADE can work with a comparatively large measurement set, and because each measurement location is illuminated by multiple \acp{mpc}. Bootstrapping with accurate location labels from JADE thus improves the performance of the \ac{nn}.



\begin{figure}[tb]
    \centering
    \subfloat[4 TXs\label{fig:heatmap4AP}]{\includegraphics[width=0.485\columnwidth]{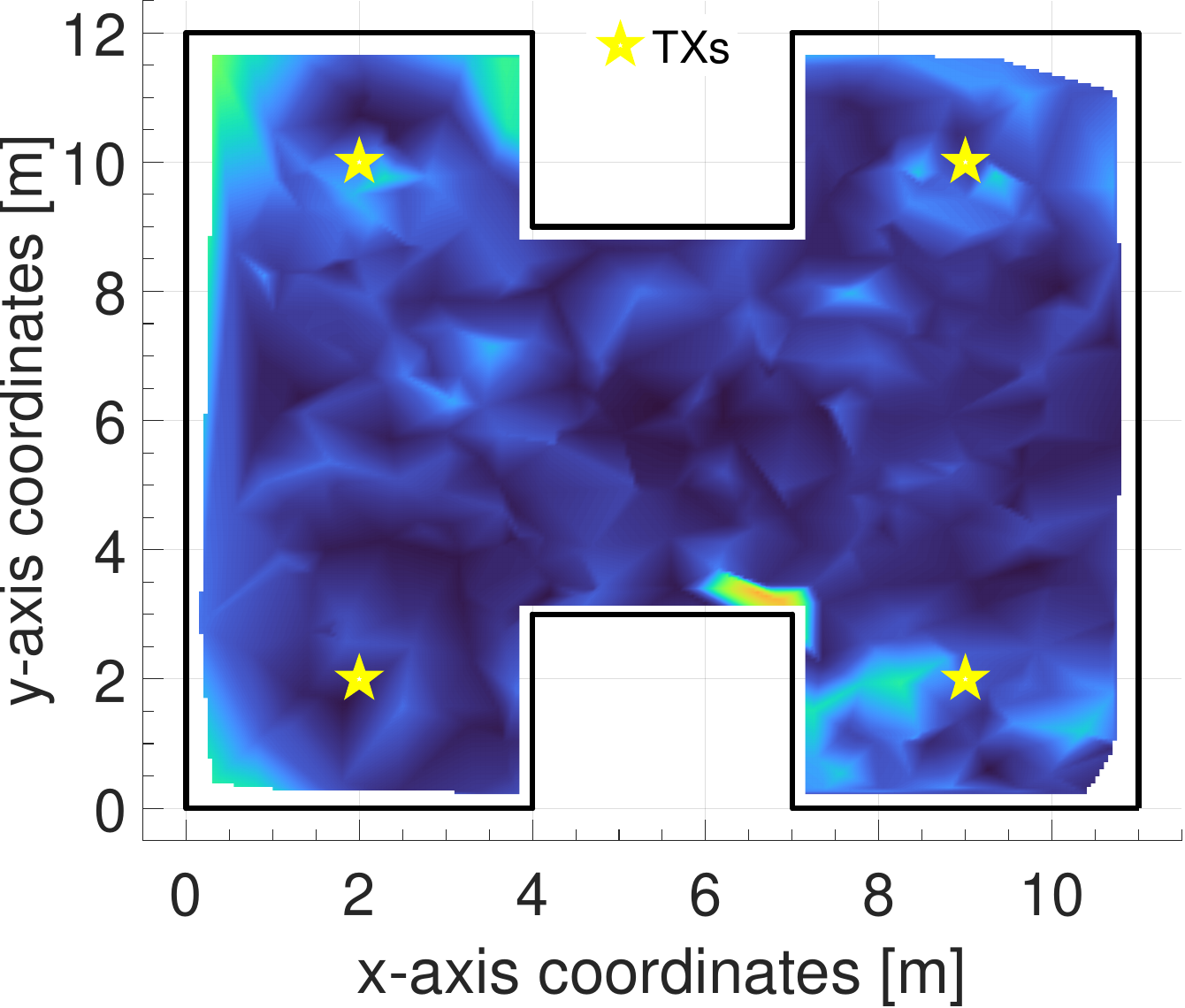}}
    \hfill
    \subfloat[3 TXs\label{fig:heatmap3AP}]{\includegraphics[width=0.4975\columnwidth]{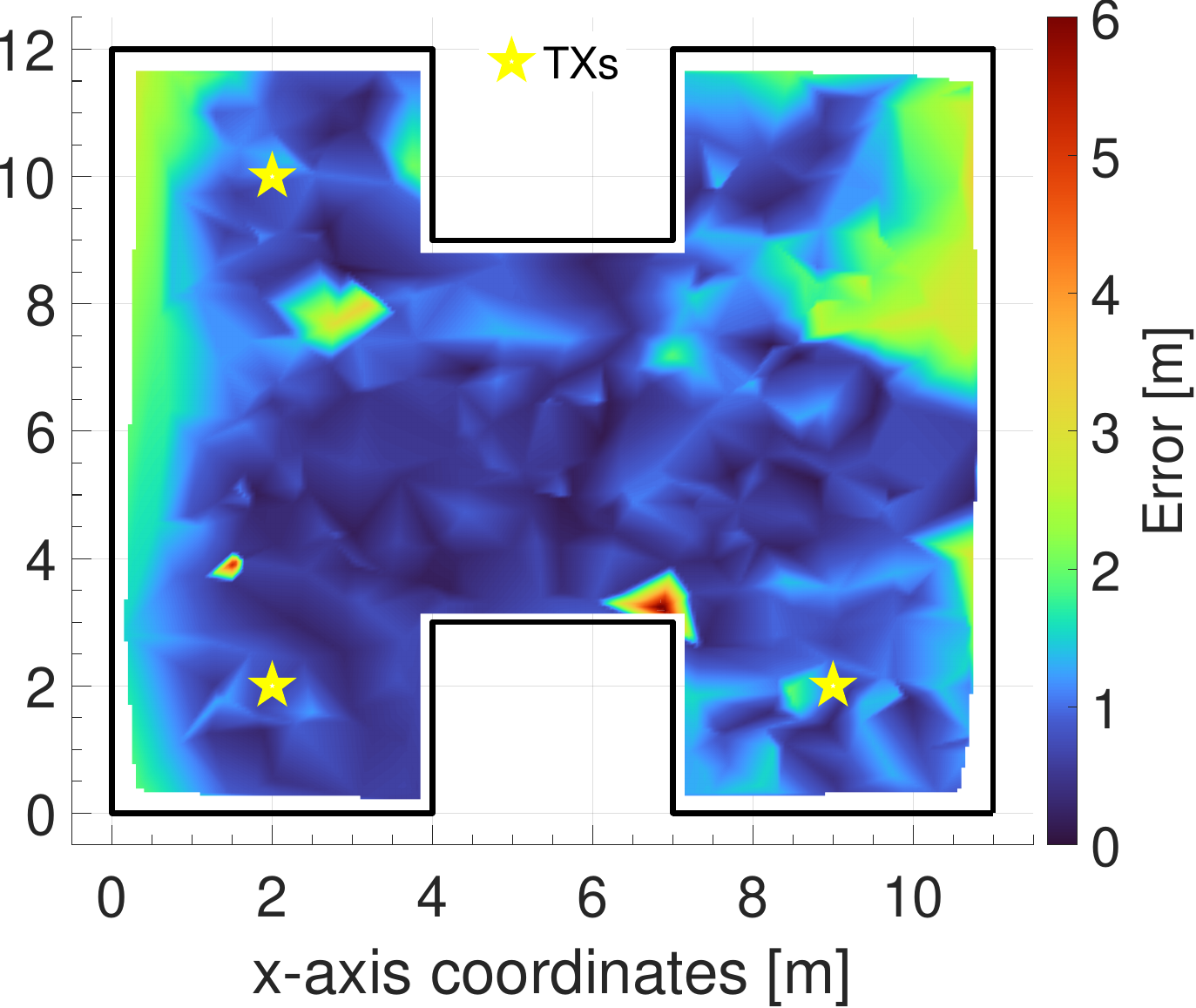}}
    \caption{Heat map of the localization errors for the H-shaped room for different numbers of transmitters, $\sigma=5^\circ$.}
    \label{fig:HroomErrorMap_numtx}
\end{figure}

Fig.~\ref{fig:HroomErrorMap} illustrates the localization error heat map in the H-shaped room, for all test locations, as estimated using the algorithm-supervised \ac{nn}. Blue hues represent low estimation errors (\textless{}1~m), light blue to green hues represent errors between 1~m and 2~m, whereas yellow to red hues convey errors between 3~m and 6~m. The results confirm that the localization error is \textless{}1~m throughout the room, except for slightly higher errors in the bottom-right section. A similar situation occurs near the top-right corner of the middle section.
The main reason is that the training dataset is not collected at uniformly distributed locations, and may thus be sparser in some areas, which also  explains why the heat map is not symmetrical.

Fig.~\ref{fig:HroomErrorMap_numtx} explores how reducing the number of transmitters affects the localization. In Fig.~\ref{fig:heatmap4AP}, we observe that removing the middle \ac{tx} reduces the illumination at the corner areas and thus increases localization errors. This includes both the \ac{los} \acp{mpc} and \ac{nlos} \acp{mpc} originating from reflections on the left and right side walls, as well as the top and bottom walls of the middle section.
Removing one additional \ac{tx}, e.g., the top right one, leads to higher localization errors in its surroundings (Fig.~\ref{fig:heatmap3AP}). However, the removal affects the whole room, as the \ac{tx} helps illuminate other room sections through the \ac{nlos} \acp{mpc} reflecting off the right wall of the room and the bottom wall of the central section.

\begin{figure}[tb]
    \centering
    \subfloat[Trained with $\sim$1400 samples\label{fig:heatmap5AP7std}]{\includegraphics[width=0.4675\columnwidth]{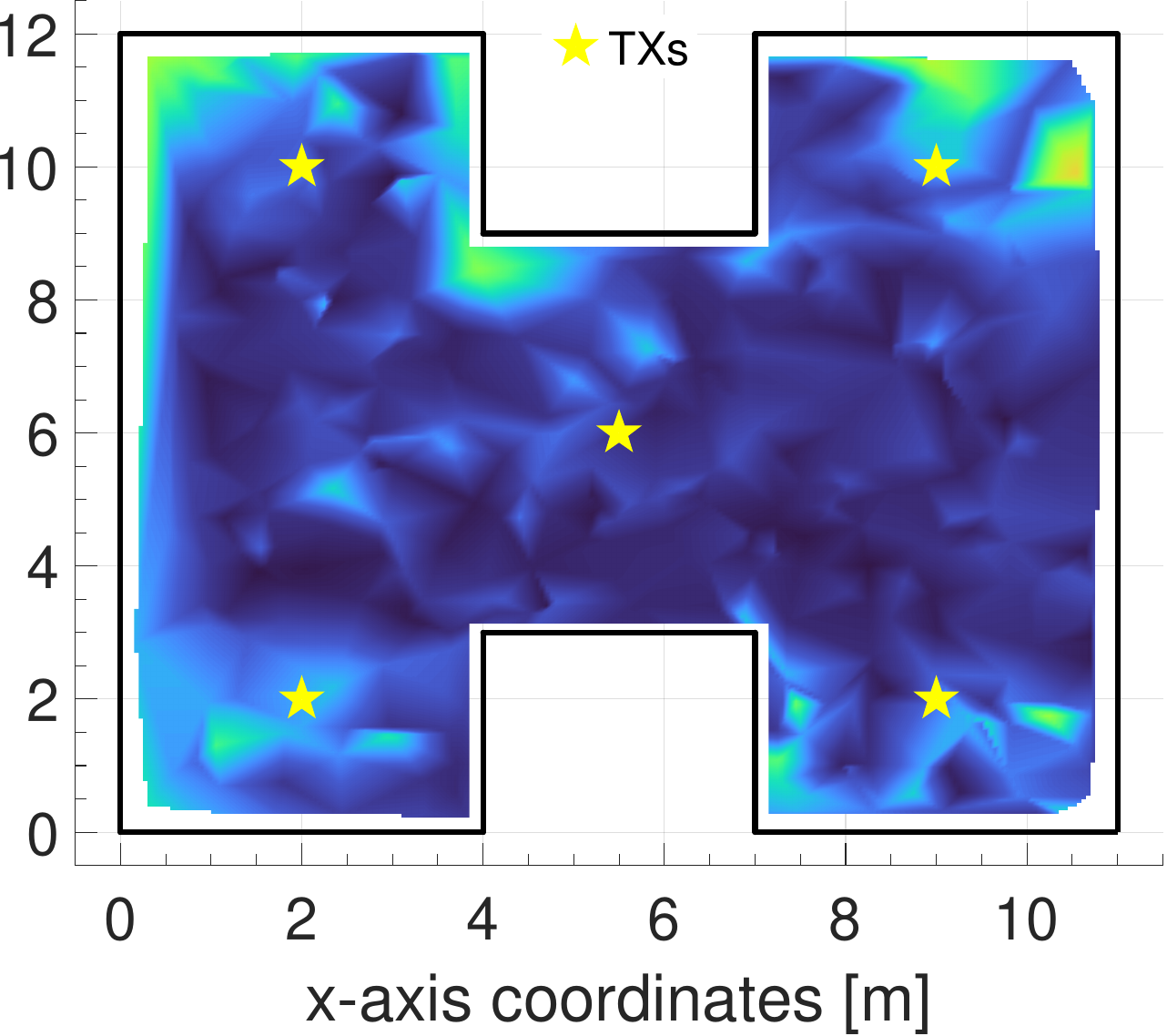}}
    \hfill
    \subfloat[Trained with $\sim$4000 samples\label{fig:heatmap5AP7std_morepts}]{\includegraphics[width=0.4975\columnwidth]{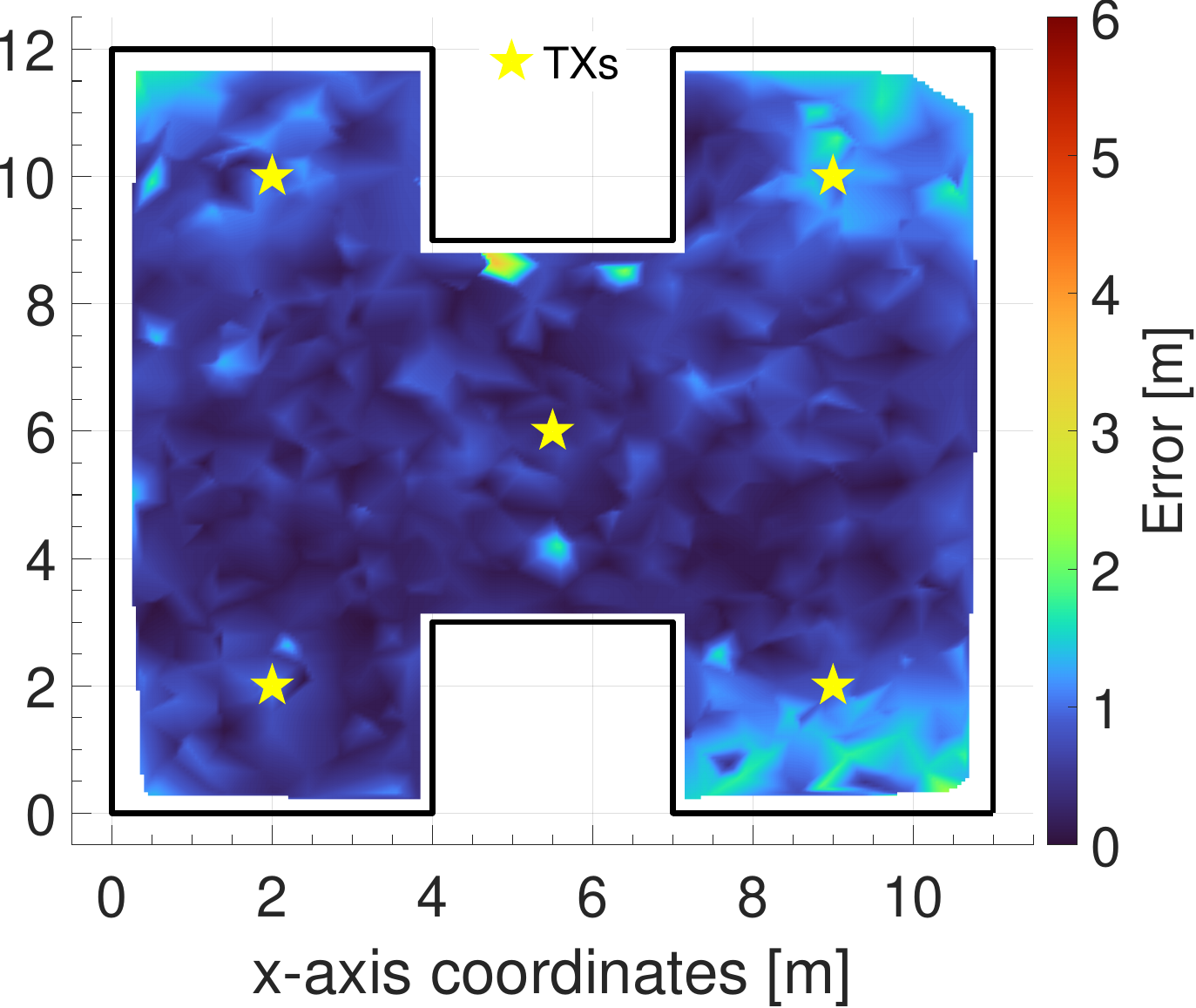}}
    \caption{Heat map of the localization errors for the H-shaped room, for AoA error standard deviation $\sigma=7^\circ$ and different NN training set sizes.}
    \label{fig:HroomErrorMap_sigma}
    \vspace{-2mm}
\end{figure}

Finally, Fig.~\ref{fig:HroomErrorMap_sigma} discusses the effect of a lower \ac{aoa} estimation accuracy, i.e., an error of $\sigma = 7^\circ$ affecting \ac{aoa} values. For this, we consider the initial room setup with 5 \acp{tx}, and evaluate the impact of training set size on localization. Fig.~\ref{fig:heatmap5AP7std} presents the same scenario as in Fig.~\ref{fig:HroomErrorMap}, with $\sigma = 7^\circ$. We observe a general error increase, where the central sections of the room are still remarkably well represented by the learned \ac{nn} model, whereas worse errors tend to concentrate at the corner sections. We stress that such a large $\sigma$ means that about 95\% of the \acp{aoa} span an interval of $\pm 14^\circ$ from the true value. Consider the case of a \ac{rx} located at any far corner of the room: a $14^\circ$ error on the \ac{aoa} of the \ac{los} path from the farthermost \ac{tx} would 
result in an offset in the \ac{rx}'s own location estimate of $\sim3$~m (measured along the circumference centered on that \ac{tx} with radius equal to the \ac{tx}--\ac{rx} distance).

We expect that the more examples the \ac{nn} sees from the bootstrapping algorithm, the higher the chances that some location estimates bear a small error.
Fig.~\ref{fig:heatmap5AP7std_morepts} shows that, in order reduce the localization error to levels comparable with Fig.~\ref{fig:HroomErrorMap}, we need to increase the training set (about 3~times the size used for Fig.~\ref{fig:heatmap5AP7std} in our scenario). With this, we achieve sub-meter errors almost everywhere, except for a few locations close to the room corners.

In general our performance evaluation shows that a tiny algorithm-supervised \ac{nn} can accurately localize a user in a complex indoor environment in the presence of two key factors: $i)$ richness of input features, coming either from \ac{los} paths, \ac{nlos} paths, or both; and $ii)$ richness of training data, so that the \ac{nn} observes sufficiently many error-prone samples.


\section{Discussion and Conclusions}
\label{sec:concl}

We presented an algorithm-supervised approach to train a tiny \ac{nn} to localize a \ac{mmw} device using \ac{adoa} measurements in an indoor environment. This relieves the burdensome task of collecting training data by harvesting location labels from a bootstrapping localization algorithm. 
We evaluated the performance of our scheme via channel measurements from a NIST \ac{mmw} channel sounder. 
The channel measurements were processed using the \ac{sage} algorithm to obtain all the \acp{mpc} at the receiver.
We emphasize that, to the best of our knowledge, this is the first algorithm that processes a full, rich, and complex indoor channel information rather than using information pre-filtered by other algorithms or device firmwares.
To identify dominant \acp{mpc} reflected off different surfaces, we proposed a recursive clustering algorithm based on \ac{dbscan}. 
The azimuthal \acp{aoa} associated with the cluster centroids were used to compute \acp{adoa}. Our performance evaluation shows that tiny \acp{nn} can accurately estimate the location of the receiver when trained with true locations. While the error inevitably increases in the algorithm-supervised case, our scheme still achieves sub-meter error in about 75\% of the cases, with a much lower computational complexity than the bootstrapping algorithm. Additional simulation results confirm that the key elements that enable high accuracy are richness of illumination from multiple \acp{mpc} and training data.

\remembertext{challenges}{Our future work will address several challenges. One key challenge is to develop schemes that can automatically decide when to switch between the bootstrapping algorithm JADE and the tiny \ac{nn} model. It is crucial to run JADE only up to the point where its computational complexity remains within acceptable limits for real-time environments. Automatically determining this optimal point will facilitate seamless transitions between the two schemes during runtime.
Additionally, we will focus on developing techniques to dynamically associate \acp{aoa} of dominant \acp{mpc} with ambient scatterers and anchors in more complex and dynamic indoor environments. Unlike our current indoor setup, dynamic scatterers in complex-shaped rooms can result in a large number of anchors. Correctly associating the \acp{adoa} in such environments will be crucial for both the bootstrapping algorithm and the \ac{nn}. Addressing these challenges will further improve robustness and accuracy in general indoor scenarios.}

\ifCLASSOPTIONcaptionsoff
  \newpage
\fi



%


\IEEEtriggercmd{\enlargethispage{-5.5cm}}
\IEEEtriggeratref{14}


\bibliographystyle{IEEEtran}
\bibliography{IEEEabrv,refLOC}  

%

\end{document}

%% file: acronyms.tex
\usepackage[acronym,shortcuts,nonumberlist,nopostdot]{glossaries}
\usepackage{glossary-mcols}

\makenoidxglossaries

\newacronym{4g}{4G}{fourth-generation}
\newacronym{5g}{5G}{fifth-generation}
\newacronym{3gpp}{3GPP}{Third-generation partnership project}
\newacronym{abft}{A-BFT}{association beamforming training}
\newacronym{abt}{ABT}{asymmetric beamforming training}
\newacronym{ad}{AD}{angle-Doppler}
\newacronym{adc}{ADC}{analog-to-digital converter}
\newacronym[firstplural=angle differences-of-arrival]{adoa}{ADoA}{angle difference-of-arrival}
\newacronym{ae}{AE}{autoencoder}
\newacronym[firstplural=angles of arrival (AoAs)]{aoa}{AoA}{angle of arrival}
\newacronym[firstplural=angles of departure (AoDs)]{aod}{AoD}{angle of departure}
\newacronym{ap}{AP}{access point}
\newacronym{api}{API}{application program interface}
\newacronym{ar}{AR}{augemented reality}
\newacronym{bi}{BI}{beacon interval}
\newacronym{brp}{BRP}{beam refinement protocol}
\newacronym{bs}{BS}{base station}
\newacronym{bti}{BTI}{beacon transmission interval}
\newacronym{cacfar}{CA-CFAR}{cell-averaging constant false alarm rate}
\newacronym{cbap}{CBAP}{contention based access period}
\newacronym{cdf}{CDF}{cumulative distribution function}
\newacronym{cir}{CIR}{channel impulse response}
\newacronym{cnn}{CNN}{convolutional NN}
\newacronym{com}{COM}{center of mass}
\newacronym{cots}{COTS}{commercial off-the-shelf}
\newacronym{csi}{CSI}{channel state information}
\newacronym{dac}{DAC}{digital-to-analog converter}
\newacronym{dbscan}{DBSCAN}{density-based spatial clustering of applications with noise}
\newacronym{dft}{DFT}{discrete Fourier transform}
\newacronym{dkf}{DKF}{discrete KF}
\newacronym{dl}{DL}{deep learning}
\newacronym{dsp}{DSP}{digital signal processor}
\newacronym{dti}{DTI}{data transmission interval}
\newacronym{ecd}{ECD}{expand-contract dilation}
\newacronym{ekf}{EKF}{extended Kalman filter}
\newacronym{em}{EM}{expectation maximization}
\newacronym{endc}{EN-DC}{enhanced UTRA-dual connectivity}
\newacronym{esprit}{ESPRIT}{estimation of signal parameters via rotational invariance techniques}
\newacronym{fcos}{FCOS}{fully connected \mbox{one-stage}}
\newacronym{fft}{FFT}{fast Fourier transform}
\newacronym{fmcw}{FMCW}{frequency-modulated continuous wave}
\newacronym{fov}{FoV}{field-of-view}
\newacronym{fpga}{FPGA}{field-programmable gate array}
\newacronym{fscn}{FSCN}{fractionally strided convolutional network}
\newacronym{ftm}{FTM}{fine time measurement}
\newacronym{gan}{GAN}{generative adversarial network}
\newacronym{gru}{GRU}{gated recurrent unit}
\newacronym{gscm}{GSCM}{geometry-based stochastic channel model}
\newacronym{hdc}{HDC}{hybrid dilated convolution}
\newacronym{hr}{HR}{heart rate}
\newacronym{hvrae}{HVRAE}{hybrid variational RNN autoencoder}
\newacronym{if}{IF}{intermediate-frequency}
\newacronym{ifft}{IFFT}{inverse FFT}
\newacronym{iir}{IIR}{infinite impulse response}
\newacronym{iot}{IoT}{Internet of things}
\newacronym{itur}{ITU-R}{International telecommunication union -- radiocommunication Sector}
\newacronym{kf}{KF}{Kalman filter}
\newacronym{knn}{K-NN}{K-nearest neighbours}

\newacronym{lm}{LM}{Levenberg-Marquardt}
\newacronym{lms}{LMS}{least mean squares}
\newacronym{los}{LoS}{line-of-sight}
\newacronym{lstm}{LSTM}{long-short term memory}
\newacronym{mac}{MAC}{medium access control}
\newacronym{mcu}{MCU}{micro-controller unit}
\newacronym{mf}{MF}{matched filter}
\newacronym{mimo}{MIMO}{multiple-input multiple-output}
\newacronym{miso}{MISO}{multiple-input single-output}
\newacronym{ml}{ML}{machine learning}
\newacronym{mmse}{MMSE}{minimum mean-square error}
\newacronym{mmw}{mmWave}{millimeter-wave}
\newacronym{mpc}{MPC}{multipath component}
\newacronym{mse}{MSE}{mean-square error}
\newacronym{music}{MUSIC}{multiple signal classification}
\newacronym{nlos}{NLoS}{non-line-of-sight}
\newacronym{nn}{NN}{neural network}
\newacronym{noma}{NOMA}{non-orthogonal multiple access}
\newacronym{ofdm}{OFDM}{orthogonal frequency-division multiplexing}
\newacronym{oks}{OKS}{object keypoints similarity}
\newacronym{pa}{PA}{power amplifier}
\newacronym{pdp}{PDP}{power-delay profile}
\newacronym{phy}{PHY}{physical layer}
\newacronym{prs}{PRS}{positioning reference signal}
\newacronym{psd}{PSD}{power spectral density}
\newacronym{pw}{PW}{pulsed wave}
\newacronym{ra}{RA}{range-azimuth}
\newacronym{rd}{RD}{range-Doppler}
\newacronym{rda}{RDA}{range-Doppler-azimuth}
\newacronym{resnet}{ResNet}{residual network}
\newacronym{rf}{RF}{radio frequency}
\newacronym{rgb}{RGB}{red-green-blue}
\newacronym{rms}{RMS}{root mean square}
\newacronym{rmse}{RMSE}{root mean square error}
\newacronym{rnn}{RNN}{recursive NN}
\newacronym{rss}{RSS}{received signal strength}
\newacronym{rssi}{RSSI}{received signal strength indicator}
\newacronym{rtt}{RTT}{round-trip time}
\newacronym{rx}{RX}{receiver}
\newacronym{rzf}{RZF}{regularized zero-forcing}
\newacronym{saf}{SAF}{spatial attention fusion}
\newacronym{sage}{SAGE}{space-alternating generalized expectation maximization}
\newacronym{sdr}{SDR}{software-defined radio}
\newacronym{slam}{SLAM}{simultaneous localization and mapping}
\newacronym{sls}{SLS}{sector-level sweep}
\newacronym{snr}{SNR}{signal-to-noise ratio}
\newacronym{soc}{SoC}{system-on-chip}
\newacronym{sp}{SP}{service period}
\newacronym{spi}{SPI}{subsample peak interpolation}
\newacronym{srs}{SRS}{sounding reference signal}
\newacronym{ssw}{SSW}{sector sweep}
\newacronym{sta}{STA}{station}
\newacronym{stft}{STFT}{short-time Fourier transform}
\newacronym{sv}{SV}{Saleh-Valenzuela}
\newacronym{svm}{SVM}{support vector machine}
\newacronym{tdoa}{TDoA}{time difference-of-arrival}
\newacronym{toa}{ToA}{time of arrival}
\newacronym{tof}{ToF}{time of flight}
\newacronym{tx}{TX}{transmitter}
\newacronym{txss}{TXSS}{transmit sweep}
\newacronym{uav}{UAV}{unmanned aerial vehicle}
\newacronym{ue}{UE}{user equipment}
\newacronym{ukf}{UKF}{unscented Kalman filter}
\newacronym{urllc}{URLLC}{ultra-reliable low-latency communications}
\newacronym{usrp}{USRP}{universal software radio peripheral}
\newacronym{v2x}{V2X}{vehicle-to-everything}
\newacronym{va}{VA}{virtual anchor}
\newacronym{vco}{VCO}{voltage-controlled oscillator}
\newacronym{vr}{VR}{virtual reality}
\newacronym{wlan}{WLAN}{wireless LAN}
\newacronym{zf}{ZF}{zero-forcing}

\setglossarystyle{alttree}
\glssetwidest{URLL-NMP}

%% file: main.bbl
\begin{thebibliography}{10}
\providecommand{\url}[1]{#1}
\csname url@samestyle\endcsname
\providecommand{\newblock}{\relax}
\providecommand{\bibinfo}[2]{#2}
\providecommand{\BIBentrySTDinterwordspacing}{\spaceskip=0pt\relax}
\providecommand{\BIBentryALTinterwordstretchfactor}{4}
\providecommand{\BIBentryALTinterwordspacing}{\spaceskip=\fontdimen2\font plus
\BIBentryALTinterwordstretchfactor\fontdimen3\font minus \fontdimen4\font\relax}
\providecommand{\BIBforeignlanguage}[2]{{%
\expandafter\ifx\csname l@#1\endcsname\relax
\typeout{** WARNING: IEEEtran.bst: No hyphenation pattern has been}%
\typeout{** loaded for the language `#1'. Using the pattern for}%
\typeout{** the default language instead.}%
\else
\language=\csname l@#1\endcsname
\fi
#2}}
\providecommand{\BIBdecl}{\relax}
\BIBdecl

\bibitem{fettweis_loc_feature_mmwave_IWCMC_2016}
F.~Lemic, J.~Martin, C.~Yarp, D.~Chan, V.~Handziski, R.~Brodersen, G.~Fettweis, A.~Wolisz, and J.~Wawrzynek, ``Localization as a feature of {mmWave} communication,'' in \emph{Proc.~IWCMC}, 2016.

\bibitem{henk2022radioLocFund}
H.~Wymeersch and G.~Seco-Granados, ``Radio localization and sensing—{P}art {I}: Fundamentals,'' \emph{{IEEE} Commun. Lett.}, 2022.

\bibitem{humanBlockage}
U.~T. Virk and K.~Haneda, ``Modeling human blockage at {5G} millimeter-wave frequencies,'' \emph{{IEEE} Trans. Antennas Propag.}, no.~3, 2020.

\bibitem{mmwaveComst2017Hanzo}
I.~A. Hemadeh, K.~Satyanarayana, M.~El-Hajjar, and L.~Hanzo, ``Millimeter-wave communications: Physical channel models, design considerations, antenna constructions, and link-budget,'' \emph{{IEEE} Commun. Surveys Tuts.}, no.~2, 2018.

\bibitem{comst2023anish}
A.~Shastri, N.~Valecha, E.~Bashirov, H.~Tataria, M.~Lentmaier, F.~Tufvesson, M.~Rossi, and P.~Casari, ``A review of millimeter wave device-based localization and device-free sensing technologies and applications,'' \emph{{IEEE} Commun. Surveys Tuts.}, no.~3, 2022.

\bibitem{ubiqPefkianakis}
S.~Sur, I.~Pefkianakis, X.~Zhang, and K.-H. Kim, ``Towards scalable and ubiquitous millimeter-wave wireless networks,'' in \emph{Proc.~ACM MobiCom}, 2018.

\bibitem{leap2019palacios}
J.~Palacios, P.~Casari, H.~Assasa, and J.~Widmer, ``{LEAP}: {L}ocation estimation and predictive handover with consumer-grade {mmWave} devices,'' in \emph{Proc.~IEEE INFOCOM}, 2019.

\bibitem{scalingMmwaveClaudio}
C.~Fiandrino, H.~Assasa, P.~Casari, and J.~Widmer, ``Scaling millimeter-wave networks to dense deployments and dynamic environments,'' \emph{Proceedings of the IEEE}, no.~4, 2019.

\bibitem{yang2021integratedLocComm}
J.~Yang, J.~Xu, X.~Li, S.~Jin, and B.~Gao, ``Integrated communication and localization in millimeter-wave systems,'' \emph{Frontiers of Information Tech.\ \& Electronic Eng.}, no.~4, 2021.

\bibitem{xiao2020overview6G}
Z.~Xiao and Y.~Zeng, ``An overview on integrated localization and communication towards {6G},'' Dec. 2021.

\bibitem{palacios2017jade}
J.~Palacios, P.~Casari, and J.~Widmer, ``{JADE}: Zero-knowledge device localization and environment mapping for millimeter wave systems,'' in \emph{Proc.~IEEE INFOCOM}, 2017.

\bibitem{accurate3D2018pefkianakis}
I.~Pefkianakis and K.-H. Kim, ``Accurate {3D} localization for 60 {GHz} networks,'' in \emph{Proc.~ACM SenSys}, 2018.

\bibitem{anish2022wcnc}
A.~Shastri, J.~Palacios, and P.~Casari, ``Millimeter wave localization with imperfect training data using shallow neural networks,'' in \emph{Proc.~IEEE WCNC}, 2022.

\bibitem{9664350}
S.~Blandino, J.~Senic, C.~Gentile, D.~Caudill, J.~Chuang, and A.~Kayani, ``Markov multi-beamtracking on 60 {GHz} mobile channel measurements,'' \emph{IEEE Open J. of Veh. Technol.}, 2022.

\bibitem{indoorLocSurvey2019zafari}
F.~{Zafari}, A.~{Gkelias}, and K.~K. {Leung}, ``A survey of indoor localization systems and technologies,'' \emph{{IEEE} Commun. Surveys Tuts.}, no.~3, 2019.

\bibitem{rssi2014icc}
M.~{Vari} and D.~{Cassioli}, ``{mmWaves} {RSSI} indoor network localization,'' in \emph{Proc.~IEEE ICC}, 2014.

\bibitem{olivier2016lightweight}
A.~Olivier, G.~Bielsa, I.~Tejado, M.~Zorzi, J.~Widmer, and P.~Casari, ``Lightweight indoor localization for 60-{GHz} millimeter wave systems,'' in \emph{Proc.~IEEE SECON}, 2016.

\bibitem{palacios2019single}
J.~Palacios, G.~Bielsa, P.~Casari, and J.~Widmer, ``Single-and multiple-access point indoor localization for millimeter-wave networks,'' \emph{{IEEE} Trans. Wireless Commun.}, no.~3, 2019.

\bibitem{blanco2022mobisys}
A.~Blanco, P.~J. Mateo, F.~Gringoli, and J.~Widmer, ``Augmenting {mmWave} localization accuracy through sub-6 {GHz} on off-the-shelf devices,'' in \emph{Proc.~ACM Mobisys}, 2022.

\bibitem{kanhere2019map}
O.~Kanhere, S.~Ju, Y.~Xing, and T.~S. Rappaport, ``Map-assisted millimeter wave localization for accurate position location,'' in \emph{Proc.~IEEE GLOBECOM}, 2019.

\bibitem{Yassin2017SimultaneousCI}
A.~Yassin, Y.~Nasser, M.~Awad, and A.~Al-Dubai, ``Simultaneous context inference and mapping using {mm-Wave} for indoor scenarios,'' in \emph{Proc.~IEEE ICC}, 2017.

\bibitem{Yassin2018GeometricAI}
A.~Yassin, Y.~Nasser, and M.~Awad, ``Geometric approach in simultaneous context inference, localization and mapping using {mm-Wave},'' in \emph{Proc.~ICT}, 2018.

\bibitem{MOSAIC2018slam}
A.~{Yassin}, Y.~{Nasser}, A.~Y. {Al-Dubai}, and M.~{Awad}, ``{MOSAIC}: Simultaneous localization and environment mapping using {mmWave} without a-priori knowledge,'' \emph{IEEE Access}, 2018.

\bibitem{clam2018Palacios}
J.~Palacios, G.~Bielsa, P.~Casari, and J.~Widmer, ``Communication-driven localization and mapping for millimeter wave networks,'' in \emph{Proc.~IEEE INFOCOM}, 2018.

\bibitem{bielsa2018indoor}
G.~Bielsa, J.~Palacios, A.~Loch, D.~Steinmetzer, P.~Casari, and J.~Widmer, ``Indoor localization using commercial off-the-shelf 60 {GHz} access points,'' in \emph{Proc.~IEEE INFOCOM}, 2018.

\bibitem{BeamAoD}
T.~T. {Tsai}, L.~H. {Shen}, C.~J. {Chiu}, and K.~T. {Feng}, ``{Beam AoD-based Indoor Positioning for 60 GHz mmWave System},'' in \emph{Proc.~IEEE VTC-Fall}, 2020.

\bibitem{Vashist2020ml}
A.~{Vashist}, D.~R. {Bhanushali}, R.~{Relyea}, C.~{Hochgraf}, A.~{Ganguly}, P.~D. {Sai Manoj}, R.~{Ptucha}, A.~{Kwasinski}, and M.~E. {Kuhl}, ``Indoor wireless localization using consumer-grade 60 {GHz} equipment with machine learning for intelligent material handling,'' in \emph{Proc.~IEEE ICCE}, 2020.

\bibitem{fing2019mitsubishi1}
M.~{Pajovic}, P.~{Wang}, T.~{Koike-Akino}, H.~{Sun}, and P.~V. {Orlik}, ``Fingerprinting-based indoor localization with commercial {mmWave WiFi} - {Part I}: {RSS} and beam indices,'' in \emph{Proc.~IEEE GLOBECOM}, 2019.

\bibitem{fing2019mitsubishi2}
P.~{Wang}, M.~{Pajovic}, T.~{Koike-Akino}, H.~{Sun}, and P.~V. {Orlik}, ``Fingerprinting-based indoor localization with commercial {mmWave} {WiFi} - {Part II}: Spatial beam {SNRs},'' in \emph{Proc.~IEEE GLOBECOM}, 2019.

\bibitem{deepL2020mitsubishi}
T.~{Koike-Akino}, P.~{Wang}, M.~{Pajovic}, H.~{Sun}, and P.~V. {Orlik}, ``Fingerprinting-based indoor localization with commercial {mmWave} {WiFi}: A deep learning approach,'' \emph{IEEE Access}, 2020.

\bibitem{wangMERLfingerprintingPart4}
P.~Wang, T.~Koike-Akino, and P.~V. Orlik, ``Fingerprinting-based indoor localization with commercial {mmWave} {WiFi}: {NLOS} propagation,'' in \emph{Proc.~IEEE GLOBECOM}, 2020.

\bibitem{wbfps2020Infocom}
P.~{Hong}, C.~{Li}, H.~{Chang}, Y.~{Hsueh}, and K.~{Wang}, ``{WBF-PS}: {WiGig} beam fingerprinting for {UAV} positioning system in {GPS}-denied environments,'' in \emph{Proc.~IEEE INFOCOM}, 2020.

\bibitem{doaLf2017}
Z.~{Wei}, Y.~{Zhao}, X.~{Liu}, and Z.~{Feng}, ``{DoA-LF}: A location fingerprint positioning algorithm with millimeter-wave,'' \emph{IEEE Access}, 2017.

\bibitem{8581503}
C.~Lai, R.~Sun, C.~Gentile, P.~B. Papazian, J.~Wang, and J.~Senic, ``Methodology for multipath-component tracking in millimeter-wave channel modeling,'' \emph{{IEEE} Trans. Antennas Propag.}, no.~3, 2019.

\bibitem{molisch2012wireless}
A.~F. Molisch, \emph{Wireless communications}.\hskip 1em plus 0.5em minus 0.4em\relax John Wiley\ \& Sons, 2012, ch.~7.

\bibitem{8319437}
C.~Gentile, P.~B. Papazian, R.~Sun, J.~Senic, and J.~Wang, ``Quasi-deterministic channel model parameters for a data center at 60 {GHz},'' \emph{{IEEE} Antennas Wireless Propag. Lett.}, no.~5, 2018.

\bibitem{ester1996density}
M.~Ester, H.-P. Kriegel, J.~Sander, X.~Xu \emph{et~al.}, ``A density-based algorithm for discovering clusters in large spatial databases with noise,'' in \emph{Proc.~AAAI KDD}, no.~34, 1996.

\bibitem{mpcTAP2018Camillo}
C.~Lai, R.~Sun, C.~Gentile, P.~B. Papazian, J.~Wang, and J.~Senic, ``Methodology for multipath-component tracking in millimeter-wave channel modeling,'' \emph{{IEEE} Trans. Antennas Propag.}, no.~3, 2019.

\bibitem{clusteringVTC2017Jian}
J.~Wang, C.~Gentile, J.~Senic, R.~Sun, P.~B. Papazian, and C.~Lai, ``Unsupervised clustering for millimeter-wave channel propagation modeling,'' in \emph{Proc.~IEEE VTC-Fall}, 2017.

\bibitem{channelParams2022Access}
H.~Tsukada, K.~Kumakura, S.~Tang, and M.~Kim, ``Millimeter-wave channel model parameters for various office environments,'' \emph{IEEE Access}, vol.~10, pp. 60\,387--60\,396, 2022.

\bibitem{7928270}
R.~Sun, P.~B. Papazian, J.~Senic, Y.~Lo, J.-K. Choi, K.~A. Remley, and C.~Gentile, ``Design and calibration of a double-directional 60 {GHz} channel sounder for multipath component tracking,'' in \emph{Proc.~EUCAP}, 2017.

\bibitem{hausmair2010sage}
K.~Hausmair, K.~Witrisal, P.~Meissner, C.~Steiner, and G.~Kail, ``{SAGE} algorithm for {UWB} channel parameter estimation,'' in \emph{COST 2100 Committee Meeting}, 2010.

\end{thebibliography}
